\pdfoutput=1
\documentclass[11pt,twoside,a4paper,cmspaper,final,collab]{cms-tdr}
\def\svnVersion{7b1d6a0}\def\svnDate{2023/07/31}\def\cmsCernNoTag{CERN-EP-2022-276}\def\cmsCernDate{\today}\def\cmsMessage{Published in the Journal of High Energy Physics as \href{http://dx.doi.org/10.1007/JHEP07(2023)210}{\doi{10.1007/JHEP07(2023)210}.}}
\begin{document}\cmsNoteHeader{EXO-21-014}

\newlength\cmsTabSkip\setlength{\cmsTabSkip}{1ex}
\newcommand{\ChiToHG}{\ensuremath{\PSGczDo\to\PH\PXXSG}\xspace}
\newcommand{\ChiToZG}{\ensuremath{\PSGczDo\to\PZ\PXXSG}\xspace}
\newcommand{\ctau}{\ensuremath{c\tau}\xspace}
\newcommand{\ctauChi}{\ensuremath{c\tau_\PSGczDo}\xspace}
\newcommand{\mChi}{\ensuremath{m_\PSGczDo}\xspace}
\newcommand{\tECAL}{\ensuremath{t_{\text{ECAL}}}\xspace}
\newcommand{\tECALi}{\ensuremath{\tECAL^{i}}\xspace}
\newcommand{\eECAL}{\ensuremath{E_{\text{ECAL}}}\xspace}
\newcommand{\eECALi}{\ensuremath{\eECAL^{i}}\xspace}
\newcommand{\NT}{\ensuremath{N_\text{TDJ}}\xspace}
\newcommand{\cbf}{\ensuremath{\mathcal{B}}\xspace}
\newcommand{\vjets}{\ensuremath{\PV}+jets\xspace}
\newcommand{\wjets}{\ensuremath{\PW}+jets\xspace}
\newcommand{\zjets}{\ensuremath{\PZ}+jets\xspace}
\newcommand{\dphimin}{\ensuremath{\Delta\phi_\text{min}}\xspace}
\newcommand{\xsecChi}{\ensuremath{\sigma_{\PSGczDo\PSGczDo}}\xspace}
\newcommand{\Ntwotagbkg}{\ensuremath{N_{\geq\text{2-tag bkg}}}\xspace}
\newcommand{\Nonetag}{\ensuremath{N_{\text{1-tag}}}\xspace}
\newcommand{\Nzerotag}{\ensuremath{N_{\text{0-tag}}}\xspace}
\newcommand{\Nkuntagged}{\ensuremath{N^k_\text{untagged}}\xspace}
\newcommand{\ebkg}{\ensuremath{\epsilon_\text{bkg}}\xspace}
\newcommand{\alphamax}{\ensuremath{\alpha_\text{max}}\xspace}
\newcommand{\betamax}{\ensuremath{\beta_\text{max}}\xspace}
\newcommand{\gammamax}{\ensuremath{\gamma_\text{max}}\xspace}
\newcommand{\statsyst}{\ensuremath{\,\text{(stat + syst)}}\xspace}
\DeclareRobustCommand{\PSGcmpDo}{{\HepParticle{\PSGc}{1}{\mp}}\Xspace}
\DeclareRobustCommand{\Pl}{{\HepParticle{\ell}{}{}}\Xspace}

\cmsNoteHeader{EXO-21-014}
\title{Search for long-lived particles using out-of-time trackless jets in proton-proton collisions at \texorpdfstring{$\sqrt{s} = 13\TeV$}{sqrt(s) = 13 TeV}}

\date{\today}

\abstract{
        A search for long-lived particles decaying in the outer regions of the CMS silicon tracker or in the calorimeters is presented.
        The search is based on a data sample of proton-proton collisions at $\sqrt{s} = 13\TeV$ recorded with the CMS detector at the LHC in 2016--2018, corresponding to an integrated luminosity of 138\fbinv. 
        A novel technique, using nearly trackless and out-of-time jet information combined in a deep neural network discriminator, is employed to identify decays of long-lived particles. 
        The results are interpreted in a simplified model of chargino-neutralino production, where the neutralino is the next-to-lightest supersymmetric particle, is long-lived, and decays to a gravitino and either a Higgs or \PZ boson. 
        This search is most sensitive to neutralino proper decay lengths of approximately 0.5\unit{m}, for which masses up to 1.18\TeV are excluded at 95\% confidence level. 
        The current search is the best result to date in the mass range from the kinematic limit imposed by the Higgs boson mass up to 1.8\TeV.
}

\hypersetup{
        pdfauthor={CMS Collaboration},
        pdftitle={Search for long-lived particles using out-of-time trackless jets in proton-proton collisions at sqrt(s) = 13 TeV},
        pdfsubject={CMS},
        pdfkeywords={CMS, SUSY, long-lived particles}
}

\maketitle

\section{Introduction}
\label{sec:intro}
Many extensions of the standard model (SM) predict the existence of neutral, weakly coupled particles with long lifetimes.
These long-lived particles (LLPs) naturally arise in supersymmetry (SUSY), \eg, in models with gauge-mediated SUSY breaking (GMSB)~\cite{GIUDICE1999419,Meade:2008wd,Buican:2008ws}, split SUSY~\cite{Giudice:2004tc,Hewett:2004nw,ArkaniHamed:2004yi,Gambino:2005eh,Arvanitaki:2012ps,ArkaniHamed:2012gw}, SUSY with weak $R$-parity violation~\cite{Fayet:1974pd,Farrar:1978xj,Weinberg:1981wj,Barbier:2004ez}, and stealth SUSY~\cite{Fan:2011yu,Fan:2012jf}.
Other SM extensions also give rise to LLPs, including scenarios with hidden valleys~\cite{Strassler:2006im,Strassler:2006ri,Han:2007ae}, baryogenesis triggered by weakly interacting massive particles~\cite{Cui:2011ab,Cui:2012jh,Cui:2014twa}, inelastic dark matter~\cite{PhysRevD.64.043502}, and twin Higgs mechanisms~\cite{Chacko:2005pe,Curtin:2015fna,Cheng:2015buv}.

In this paper, we describe a search that focuses on LLPs that decay in the outer regions of the CMS tracker or within the calorimeters forming a signature with jets that are nearly trackless and out-of-time (delayed) with respect to the LHC collisions, hereafter referred to as TD jets.
This search uses a sample of proton-proton ($\Pp\Pp$) collisions collected in 2016--2018 at $\sqrt{s} = 13\TeV$ with the CMS detector at the CERN LHC, corresponding to an integrated luminosity of 138\fbinv.
Jets produced by LLP decays that occur in the calorimeters or in the last few layers of the silicon tracker will have low track multiplicity compared to prompt jets that originate directly from the primary collision point.
For simplicity, we will refer to such signal jets as trackless, though they may only be nearly so.
Additionally, these signal jets will exhibit significant time delays with respect to those produced by SM backgrounds.
The time delay arises from two different causes.
The first cause, which dominates for LLP masses below about 400\GeV, is that LLP decays occur at macroscopic distances from the primary collision point, resulting in decay products that traverse a longer path to reach the calorimeters.
The second cause, which dominates for LLP masses above about 600\GeV, is that high-mass LLPs travel measurably slower than the speed of light.
The properties of the tracks and electromagnetic calorimeter (ECAL) hits associated with the jets are combined in a deep neural network (DNN) discriminator, which is used to distinguish between jets from LLP decays and those from SM backgrounds.

The ATLAS and CMS Collaborations have previously searched for LLPs using jets containing displaced tracks~\cite{CMS:2020iwv,ATLAS:2017tny,ATLAS:2018pvw,ATLAS:2022zhj,ATLAS:2021jig}.
These searches rely on tracking information and thus yield excellent sensitivity to LLPs with proper decay lengths (\ctau), where $\tau$ is the lifetime of the LLP in its rest frame, below $\approx$0.2\unit{m} when the LLPs predominantly decay in the silicon trackers.
Conversely, searches using displaced vertices in the ATLAS muon spectrometer~\cite{ATLAS:2022gbw} and a search by the CMS experiment using localized regions with high hit multiplicity in the muon system~\cite{CMS:2021juv} result in excellent sensitivity to LLPs with \ctau above 1\unit{m}.
The search reported in this paper covers the intermediate \ctau range, 0.3--1.0\unit{m}, and extends a previous CMS search using delayed jets~\cite{CMS:2019qjk} down to the kinematic limit imposed by the Higgs mass through the use of a dedicated TD jet tagger that makes optimal use of the time delay and trackless features of the jet.

\begin{figure}[htbp]
        \centering
        \includegraphics[width=0.3\textwidth]{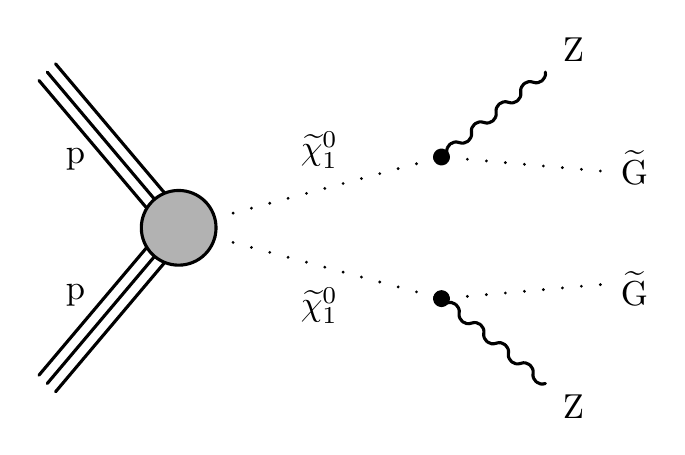}
        \includegraphics[width=0.3\textwidth]{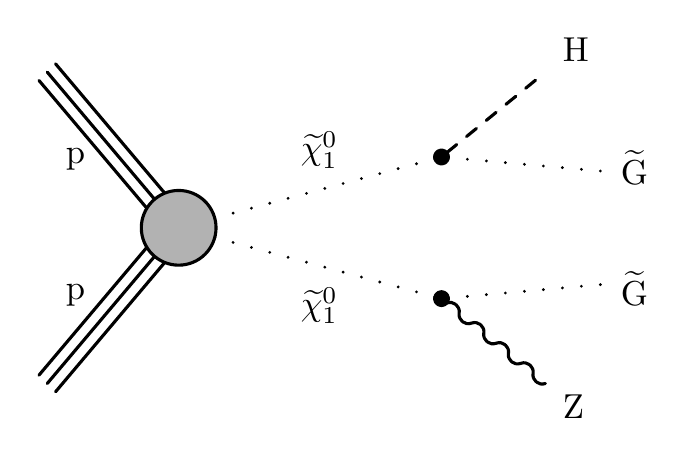}
        \includegraphics[width=0.3\textwidth]{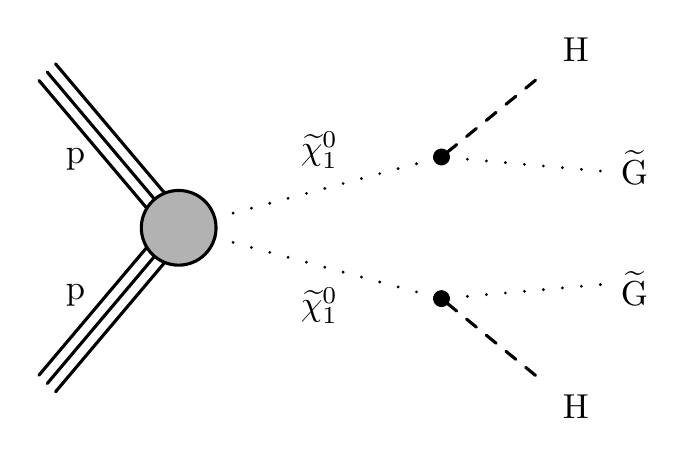}
        \caption{
                Feynman diagrams of the effective neutralino pair production in the GMSB simplified model in which the two neutralinos decay into two gravitinos (\PXXSG) and two \PZ bosons (left), a \PZ and a Higgs boson (\PH) (center), or two Higgs bosons (right).
        }
        \label{fig:feynmanDiagrams}
\end{figure}

While this search is sensitive to many models predicting LLPs, we interpret its results using a simplified model of GMSB chargino-neutralino production~\cite{Dimopoulos:1996vz,Matchev:1999ft}.
The simplified model describes Higgsino-like charginos (\PSGcpmDo) and neutralinos (\PSGczDo, \PSGczDt).
All of them are assumed to be nearly degenerate in mass, with \PSGczDo being the lightest of them.
They are produced in pairs in the following combinations: $\PSGczDo\PSGczDt$, $\PSGczDo\PSGcpmDo$, $\PSGczDt\PSGcpmDo$, and $\PSGcpmDo\PSGcmpDo$.
Because of the near mass-degeneracy, both \PSGczDt and \PSGcpmDo decay to \PSGczDo and to other low transverse momentum (\pt) particles, leading to effective \PSGczDo pair production.
Each \PSGczDo will subsequently decay to a Higgs or \PZ boson and a gravitino (\PXXSG), which is the lightest SUSY particle, thus conserving $R$-parity.
The search focuses on hadronic decays of the Higgs or \PZ boson.
Signal events typically have large missing transverse momentum resulting from the momentum of the gravitinos, which do not interact with the detector, or from the Lorentz boost of the neutralino when the neutralino decays outside of the calorimeter volume.
Feynman diagrams of the effective neutralino pair production in the simplified model are shown in Fig.~\ref{fig:feynmanDiagrams}.

The rest of this paper is organized as follows.
A brief description of the CMS detector is given in Section~\ref{sec:detector}.
Section~\ref{sec:simulation} provides a summary of the various simulated samples used in the analysis.
The event reconstruction and selection are described in Section~\ref{sec:evtsel}, while the DNN-based TD jet discriminator is discussed in Section~\ref{sec:tagger}.
The background estimation and systematic uncertainties are detailed in Sections~\ref{sec:bkg}~and~\ref{sec:systematics}, respectively.
We report and interpret the results in Section~\ref{sec:results}.
Finally, a summary is given in Section~\ref{sec:summary}.
The detailed results of this paper are found in the associated HEPData record~\cite{hepdata}.

\section{The CMS detector}
\label{sec:detector}
The central feature of the CMS detector is a superconducting solenoid of 6\unit{m} internal diameter, providing a magnetic field of 3.8\unit{T}.
Within the solenoid volume are a silicon pixel and strip tracker, a lead tungstate crystal ECAL, and a brass and scintillator hadron calorimeter (HCAL), each composed of a barrel and two endcap sections.

The ECAL consists of 75\,848 lead tungstate crystals, which cover $\abs{\eta} < 1.48 $ in the barrel region and $1.48 < \abs{\eta} < 3.00$ in the two endcap regions.
The excellent signal-to-noise ratio and stable pulse shape of the ECAL sensors allow for timing measurements with the best resolution achievable at very large energies of better than 100\unit{ps} per hit~\cite{CMS:2019zxa}.
The HCAL is composed of cells of width 0.087 in $\eta$ and azimuth ($\phi$, in radians) for the region $\abs{\eta} < 1.74$, progressively increasing to a maximum of 0.174 for larger values of $\abs{\eta}$, along with the forward calorimeters extending the $\eta$ coverage provided by the barrel and endcap detectors.
The outer tracker layers extend to a radial distance of 1.16\unit{m} in the barrel; the front face of the ECAL crystals are located at a radial distance of 1.29~\unit{m}, and the crystals extend radially for 23~\unit{cm}; and the HCAL covers the radial distances between 1.77~and~2.95~\unit{m}.
Muons are measured in the range $\abs{\eta} < 2.4$, with detection planes embedded in the steel flux-return yoke outside the solenoid and made using three technologies: drift tubes (DTs), cathode strip chambers (CSCs), and resistive-plate chambers.

Events of interest are selected using a two-tiered trigger system.
The first level, composed of custom hardware processors, uses information from the calorimeters and muon detectors to select events at a rate of around 100\unit{kHz} within a fixed latency of about 4\mus~\cite{CMS:2020cmk}.
The second level, known as the high-level trigger, consists of a farm of processors running a version of the full event reconstruction software optimized for fast processing, and reduces the event rate to around 1\unit{kHz} before data storage~\cite{CMS:2016ngn}.
A more detailed description of the CMS detector, together with a definition of the coordinate system used and the relevant kinematic variables, can be found in Ref.~\cite{CMS:2008xjf}.

\section{Simulated samples}
\label{sec:simulation}

Several simulated event samples are used to model the signal and SM background processes, and as training samples for the TD jet discriminator.
Samples of top quark-antiquark pair (\ttbar) and single top quark events, \wjets and \zjets (collectively called \vjets) events, and events composed uniquely of jets produced through the strong interaction, referred to as quantum chromodynamics (QCD) multijet events, are produced with the Monte Carlo (MC) generator \MGvATNLO~2.2.2~\cite{Alwall:2014hca} at leading-order (LO).
Up to three and four additional partons are considered in the matrix element (ME) calculations for the generation of the \vjets and top quark samples, respectively.
The MLM jet matching scheme~\cite{Alwall:2007fs,Alwall:2008qv} is used.
The background samples are normalized according to the most precise cross section calculations available, typically corresponding to next-to-LO (NLO) or next-to-NLO (NNLO) precision~\cite{Gavin:2010az, Gavin:2012sy, Czakon:2011xx, Alwall:2014hca, Alioli:2009je}.

{
\tolerance=1600
The SUSY signal samples are generated at LO in perturbative QCD using the \MGvATNLO~2.4.2 generator~\cite{Alwall:2014hca} with up to two additional partons in the final state at the ME level, and using the MLM jet matching scheme.
The corresponding cross sections for effective $\PSGczDo\PSGczDo$ production, assuming nearly mass-degenerate Higgsino-like $\PSGcpmDo$, $\PSGczDt$, and $\PSGczDo$, are computed to NLO precision including next-to-leading logarithmic corrections~\cite{NLONLL1,NLONLL2,NLONLL3,NLONLL4,NLONLL5,Borschensky:2014cia,Fuks:2013vua,Fuks:2012qx}.
In the calculation, all SUSY particles other than $\PSGcpmDo$, $\PSGczDt$, $\PSGczDo$, and $\PXXSG$ are assumed to be too heavy to participate in the interaction.
We follow the convention of real mixing matrices and signed masses for the neutralinos and charginos~\cite{Skands:2003cj} used in previous CMS searches~\cite{CMS:2018szt,CMS:2017bki}.
\par
}

For the simulated samples corresponding to the 2016 data-taking period, the NNPDF~v3.0 LO and NLO parton distribution functions (PDFs)~\cite{Ball:2014uwa,Ball:2017nwa} are used for samples generated at LO and NLO, respectively.
For the simulated samples corresponding to the 2017--2018 data-taking periods, the NNPDF~v3.1 NNLO PDFs are used.
The simulated events at the ME level are interfaced with \PYTHIA8.226~\cite{Sjostrand:2014zea} or a later version to simulate the shower and hadronization of partons, and the underlying event description.
The CUETP8M1~\cite{Khachatryan:2015pea} and CP5~\cite{Sirunyan:2019dfx} tunes are used for the simulated samples corresponding to the 2016 and 2017--2018 data-taking periods, respectively.

For all simulated samples, the detector response is modeled using a detailed description of the CMS detector based on \GEANTfour~\cite{GEANT4}.
Object and event reconstruction are performed with the same algorithms as are used for the CMS data.
Minimum bias events are superimposed on each simulated hard scattering event to reproduce the effect of extra $\Pp\Pp$ interactions within the same or neighboring bunch crossings as the recorded event (pileup).
Events are weighted such that the distribution of the number of interactions per bunch crossing agrees with that observed during each data-taking period.

\section{Event reconstruction and selection}
\label{sec:evtsel}

The particle-flow (PF) algorithm~\cite{CMS:2017yfk} aims to reconstruct and identify each individual particle in an event as an electron, photon, muon, charged or neutral hadron, with an optimized combination of information from the various elements of the CMS detector.
These reconstructed particles are referred to as PF candidates.
Photons are identified as ECAL energy clusters not linked to the extrapolation of any charged particle trajectory to the ECAL~\cite{CMS:2015myp}.
The energy of photons is obtained from the ECAL measurement.
Electrons are identified as a primary charged particle track and potentially many ECAL energy clusters corresponding to this track extrapolation to the ECAL and to possible bremsstrahlung photons emitted along the way through the tracker material~\cite{CMS:2020uim,CMS-DP-2020-021}.
The electron momentum is estimated by combining the energy measurement in the ECAL with the momentum measurement in the tracker.
Muons are identified as tracks in the central tracker consistent with either a track or several hits in the muon system, and associated with calorimeter deposits compatible with the muon hypothesis.
Hadronically decaying \PGt leptons are reconstructed with the hadron-plus-strips (HPS) algorithm~\cite{CMS:2015pac,CMS:2018jrd} on the basis of the number of tracks and of the number of ECAL strips in the $\eta$-$\phi$ plane with energy deposits, in the 1-prong, 1-prong + \Pgpz, and 3-prong decay modes. 
A cut-based selection based on isolation sums~\cite{CMS:2018jrd} is used to suppress the rate for quark- and gluon-initiated jets to be misidentified as hadronic \PGt candidates. 

Charged hadrons are identified as charged particle tracks neither identified as electrons, nor as muons.
The energy of charged hadrons is determined from a combination of the track momentum and the corresponding ECAL and HCAL energies, corrected for the response function of the calorimeters to hadronic showers.
Finally, neutral hadrons are identified as HCAL energy clusters not linked to any charged hadron trajectory, or as a combined ECAL and HCAL energy excess with respect to the expected charged hadron energy deposit.
The energy of neutral hadrons is obtained from the corresponding corrected ECAL and HCAL energies.

For each event, hadronic jets are clustered from PF candidates using the infrared- and collinear-safe anti-\kt algorithm~\cite{Cacciari:2008gp, Cacciari:2011ma} with a distance parameter of 0.4.
The jet momentum is determined as the vector sum of all particle momenta in the jet, and is found from simulation to be, on average, within 5--10\% of the true momentum over the entire \pt spectrum and detector acceptance.
Although charged particles resulting from LLP decays that occur farther from the interaction point may be misidentified as neutral PF candidates, their contribution to the jet energy scale and resolution is correctly accounted for on average.
Pileup interactions can produce additional tracks and calorimetric energy depositions, increasing the apparent jet momentum.
To mitigate this effect, tracks consistent with pileup vertices are discarded and an offset correction is applied to correct for remaining contributions.
We have verified that this pileup mitigation procedure does not reduce the signal efficiency.
Jet energy corrections are derived from simulation studies so that the average measured energy of jets becomes identical to that of particle level jets.
In situ measurements of the momentum balance in dijet, photon+jet, {\PZ}+jet, and multijet events are used to determine any residual differences between the jet energy scale in data and in simulation, and appropriate corrections are made~\cite{CMS:2016lmd}.
Using {\PZ}+jet events, we verified that these jet energy corrections, as well as their uncertainties, remain accurate when applied to jets with high scores of the TD jet tagger.
The jet energy resolution amounts typically to 15--20\% at 30\GeV, 10\% at 100\GeV, and 5\% at 1\TeV~\cite{CMS:2016lmd}.

The vector \ptvecmiss is computed as the negative vector \pt sum of all the PF candidates in an event, and its magnitude is denoted as \ptmiss~\cite{CMS:2019ctu}.
The primary vertex (PV) is taken to be the vertex corresponding to the hardest scattering in the event, evaluated using tracking information alone, as described in Section 9.4.1 of Ref.~\cite{CMS-TDR-15-02}.

\subsection{Jet time reconstruction}
\label{sec:jettime}

Jets from signal events arrive at the ECAL later than particles produced at the PV by 1--10\unit{ns}, depending on the LLP mass.
Therefore, measuring the jet time delay with respect to a particle produced at the PV and traveling at the speed of light helps to discriminate between signal and background.
The time of arrival of a jet at the ECAL, \tECAL, is calculated based on an energy-weighted sum of the arrival times reconstructed from the signal pulses in each ECAL crystal associated with the jet
\begin{equation}
        \label{eqn:jetTime}
        \tECAL = \frac{\sum^{N_\text{crystal}}_{i=1}{\tECALi\eECALi}}{\sum^{N_\text{crystal}}_{i=1}{\eECALi}},
\end{equation}
where \tECALi and \eECALi are the timestamp and reconstructed energy of the signal pulse in crystal $i$, respectively, and $N_\text{crystal}$ is the number of crystals associated with the jet.
Details of the ECAL time reconstruction algorithm and performance can be found in Ref.~\cite{Chatrchyan:2009aj}.
Accounting for clock jitter, collision beam spot size, and time-dependent calibration effects, the effective jet time resolution is $\sim 400$--600\unit{ps} for jets with \pt ranging from 30--150\GeV.
During collisions, there are variations in the clock signal distribution between different regions of the ECAL and different LHC fills, which are corrected using special calibration data, but these corrections may be imperfect.
As a result, there may be a degradation in the time resolution for jets~\cite{delRe:2015hla}.
We validate and correct, where necessary, the simulated prediction of the jet time mean and resolution using a pure sample of \PQb-tagged jets from dilepton \ttbar events.
The high purity of the \PQb-tagged jets from the dilepton \ttbar sample ensures that the jet time measurement is not contaminated by pileup jets, which have a different timing response.
We apply a Gaussian smearing procedure~\cite{CMS:2016lmd}: the jet time in the \ttbar simulation sample is multiplied by a random number drawn from a Gaussian distribution.
We determine the Gaussian mean and standard deviation parameters that minimize the $\chi^{2}$ test statistic~\cite{CowanPDGStat} calculated between the data and the simulation-based prediction.
The same jet time correction is also applied to the signal simulation, and the correction uncertainties are propagated to the TD jet tagger efficiency.
The systematic uncertainties in the signal yield resulting from the time correction range from 3 to 18\%, generally increasing with decreasing \PSGczDo mass. 

\subsection{Event selection}
\label{sec:evt_sel}
Events for this search are collected with a trigger requiring $\ptmiss > 120\GeV$~\cite{CMS:2016ngn} when reconstructed online.
We define a preselection by requiring events to have $\ptmiss > 200\GeV$ when reconstructed offline, which ensures a trigger selection efficiency above 99\%.
Background events can occur because of mismeasurement of high-\pt jets, which results in the alignment of the $\phi$ coordinates of \ptvecmiss and the nearby mismeasured jet.
The \dphimin is defined as the $\phi$ angle between \ptvecmiss and the closest jet with $\pt > 30\GeV$.
For the QCD multijet background, the \dphimin distribution is predominantly near zero, while for signal events the distribution is more uniform because the \ptmiss results from the \PXXSG escaping the detector rather than any of the jets being mismeasured.
We select events with $\dphimin > 0.5$, reducing the QCD multijet background by one order of magnitude while retaining 70--85\% of the signal events.

The main tool used in this analysis to differentiate between signal and background is the TD jet tagger, discussed in Section~\ref{sec:tagger}.
We consider jets with $\pt > 30\GeV$ and $\abs{\eta}<1.0$, and count the number of jets that pass the TD jet tagger selection defined in Section~\ref{sec:tagger}.
Jets in the barrel region of the detector with $\abs{\eta}$ between 1.00--1.48 are not considered because the rate of background jets passing the TD jet tagger selection is significantly larger due to higher tracking inefficiency in this region.
To ensure that jets primarily composed of nonisolated muons, electrons, or photons are not misidentified as TD-tagged jets, we veto any jet with the energy fractions of these particles larger than 0.6, 0.6, or 0.8, respectively.
In addition, jets that are within $\Delta R = \sqrt{\smash[b]{(\Delta\phi)^{2}+(\Delta\eta)^{2}}} = 0.4$ of any identified and isolated muon, electron, or photon are also not considered.

To suppress background from \wjets and \ttbar, we reject any event with an identified and isolated muon, electron, or hadronically decaying tau candidate. 
The number of TD-tagged jets ($\NT$) is used to define the different data samples used in the analysis.
The signal region (SR) is defined as one comprising events with $\NT\geq 2$. 
To estimate the background in the SR, a background-enriched control region (CR) that comprises events with exactly one TD-tagged jet is used.
Events with zero TD-tagged jets are also expected to arise primarily from background processes and are used to validate the background estimation method, as described in Section~\ref{sec:bkg}.
Because of the large background suppression factor exceeding $10^{3}$ from the presence of each additional TD-tagged jet, the impact of signal contamination in the zero and one TD-tagged jet region on the analysis sensitivity was found to be less than 1\%.

\subsection{Cosmic ray muon and beam halo vetoes}
\label{sec:CosmicsAndBeamHalo}

Noncollision backgrounds from cosmic ray muons or beam halo particles are typically trackless because their trajectories do not coincide with the luminous region known as the beam spot, which the track reconstruction algorithm assumes to be the origin of particle trajectories~\cite{CMS:2014pgm}.
Because we require at least two TD-tagged jets, these noncollision backgrounds only enter the SR if the particles are oriented in specific ways.
Given the specific trajectory configurations required, such backgrounds are rare.
To suppress them to negligible levels, we use particular features of such events to veto them.

Collisions between beam protons and an upstream collimator can result in beam halo particles that pass through the calorimeter volume parallel to the beam axis.
These beam halo particles may induce localized energy deposits in the calorimeters causing delayed signals that are misreconstructed as multiple jets with nearly equal $\phi$ coordinates~\cite{CMS:2019ctu}.
In addition, such jets tend to have a very small number of ECAL hits.
Therefore, we veto events in which the two TD-tagged jets have $\abs{\Delta\phi} < 0.05$ and at least one of the two jets has 10 or fewer ECAL hits.
This veto has a signal efficiency between 98 and 100\% depending on the signal mass and \ctau.
Events are also rejected by the so-called CSC beam halo filter~\cite{CMS:2019ctu} if they contain reconstructed tracks parallel to the beam axis using segments in the CSCs.

Events containing a cosmic ray muon can enter the SR if the muon traverses the calorimeter volume without passing through the tracker, producing localized energy deposits that can give rise to multiple apparent jets that are delayed in time.
Cosmic ray muon background events can be identified by the presence of track segments in the DT muon detector planes, which lie on a line with the calorimeter hits comprising the jets.
Segments are clustered with a density-based algorithm~\cite{DBSCANcluster} and a minimum distance parameter of 1.4\unit{m}, and segments from at least three different sets of DT muon detector planes are required in each cluster.
If two or more DT segment clusters are found, the geometric centroid positions of the two clusters closest to the calorimeter hits of the jets in the $z$ coordinate are used to define a line.
If the distance of closest approach from the centroid of the ECAL hits to the line is less than 0.5\unit{m}, then the event is interpreted as a cosmic ray muon event and rejected from the SR and CRs. This veto has a signal efficiency larger than 99.9\%.

\section{Trackless delayed jet tagger}
\label{sec:tagger}

Jets produced by LLPs that decay in the outer region of the CMS tracking volume or inside the calorimeters tend to be trackless and delayed in time, features that are not exhibited by SM backgrounds.
We trained a DNN to optimally exploit these features for discriminating between TD and SM jets.

The TD jet tagger is based on information from tracks, PF candidates, and calorimeter hits associated with the jet, expressed as 22 input features.
The tracking-related observables we use include the number of and scalar \pt sum of tracks in the jet, the scalar \pt sum of tracks matched to the PV in the jet, the minimum $\Delta R$ values between the jet axis and any track as well as between the jet axis and any track associated with the PV, and three variables sensitive to pileup: \alphamax, \betamax, and \gammamax.
These variables are defined as
\begin{equation}
        \alphamax  = \frac{ \max_{\text{trk} \in \text{jet} \cap \text{PV}} \pt^\text{trk}}{ \sum_{\text{trk}\in\text{jet}} \pt^\text{trk}},  
        \betamax   = \frac{ \max_{\text{trk} \in \text{jet} \cap \text{PV}} \pt^\text{trk}}{\pt^\text{jet}},                                  
        \gammamax  = \frac{ \max_{\text{trk} \in \text{jet} \cap \text{PV}} \pt^\text{trk}}{E^{\text{jet}}},
\end{equation}
where $\text{trk} \in \text{jet} \cap \text{PV}$ denotes the track that belongs to the jet and is associated with the PV, $\text{trk} \in \text{jet}$ denotes the track that belongs to the jet, $\pt^\text{trk}$ is the track \pt, and $\pt^\text{jet}$ and $E^\text{jet}$ are the jet \pt and energy, respectively.
The energy fractions associated with hadrons, electrons, and photons, and the number of PF candidate constituents of the jet are also used as inputs to the TD jet tagger.
Finally, several variables calculated using ECAL hits associated with the jet are used: the number of ECAL hits in the jet, the ratio of the energy sum of all ECAL hits in the jet to $E^\text{jet}$, the jet time calculated as in Eq.~(\ref{eqn:jetTime}), the semi-major and semi-minor axes of the ECAL hits in the jet~\cite{CMS:2019zxa}, and the fragmentation function.
The latter is a function that describes how the ECAL hits contribute to the jet energy considering their multiplicity~\cite{CMS-PAS-JME-13-002}.
Jets from signal LLP decays tend to have low values of \alphamax, \betamax, \gammamax, charged energy fraction, semi-major and semi-minor axes, and the fragmentation function, and large values of $\Delta R$ between the jet axis and tracks.
The charged and neutral energy fractions are lower and higher, respectively, for signal jets than for background jets not because the jet composition is different, but rather because the PF algorithm misidentifies the particle types because of reduced tracking efficiency for charged particles produced at large displacement.

Simulation samples are used to train and evaluate the performance of the DNN.
The signal sample consists of GMSB signal events with a \PSGczDo mass of 400\GeV and \ctau of 1\unit{m}.
These events are used only for training or evaluating the performance of the DNN and are not used to predict the signal yields for the final statistical analysis.
This \PSGczDo mass was chosen for the training sample because our search targets the low-mass range starting from the kinematic limit imposed by the Higgs boson mass, where previous searches have limited sensitivity.
In signal events, the Higgs or \PZ boson, produced by the decay of the LLP, subsequently decays to a pair of partons that can each be reconstructed as a displaced jet.
Such jets are labeled as signal jets if they geometrically coincide with a parton that originates from LLP decays that occur between the last layers of the tracker and the outer surface of the HCAL (decay radii between 0.30 and 1.84\unit{m}).
Jets are required to have $\abs{\eta}<1.48$ to be included in the training sample. 
This selection is further restricted to $\abs{\eta}<1$ in the search to improve the final signal-to-background ratio as the background increases in the higher $\eta$ range because of tracking inefficiencies.
Background samples consist of a mixture of QCD multijet, \wjets, $\PZ(\PGn\PAGn)$+jets, and \ttbar events weighted in proportion to their production cross sections.
The signal and background samples obtained after applying the selections described in Section~\ref{sec:evt_sel} contain 0.3 and 2.2 million events, respectively. 
The samples are split into training, validation, and testing sets, accounting for 40, 10, and 50\% of the full set, respectively.
The validation set is used to evaluate the training process and monitor for overfitting, while the testing set is used to measure the final performance after training.
We show comparisons between the simulation prediction and the observed data distributions for a few key input variables to the TD jet tagger in Appendix~\ref{TDJInputs}, confirming that the simulation provides an accurate description.  

A fully connected DNN architecture was developed with machine learning software frameworks \textsc{Keras}~\cite{chollet2015keras} and \textsc{TensorFlow}~\cite{tensorflow2015-whitepaper}.
The network architecture consists of four fully connected hidden layers, with 64, 32, 16, and 8 nodes, respectively, and an output layer with two nodes, corresponding to the two classes, signal and SM background.
Rectified linear unit~\cite{relu1,relu2} activation functions are used in the hidden layers, while a softmax activation function~\cite{softmax1,softmax2} is applied to the output layer, and a 10\% dropout rate~\cite{dropout} is applied to all layers to prevent overfitting.
We use a batch size of 512, categorical cross entropy as the loss function, and the Adam optimizer~\cite{adam} with an initial learning rate of $10^{-3}$.
Training is performed for up to 1\,000 epochs, or cycles, over the full training set.
Early stopping is applied to mitigate overfitting: after each epoch, the loss function is evaluated on the validation set and the training process is interrupted if after 100 successive epochs, the validation loss has not decreased.
The model corresponding to the smallest validation loss is used.

\begin{figure}[htpb]
        \centering
        \includegraphics[width=0.49\textwidth]{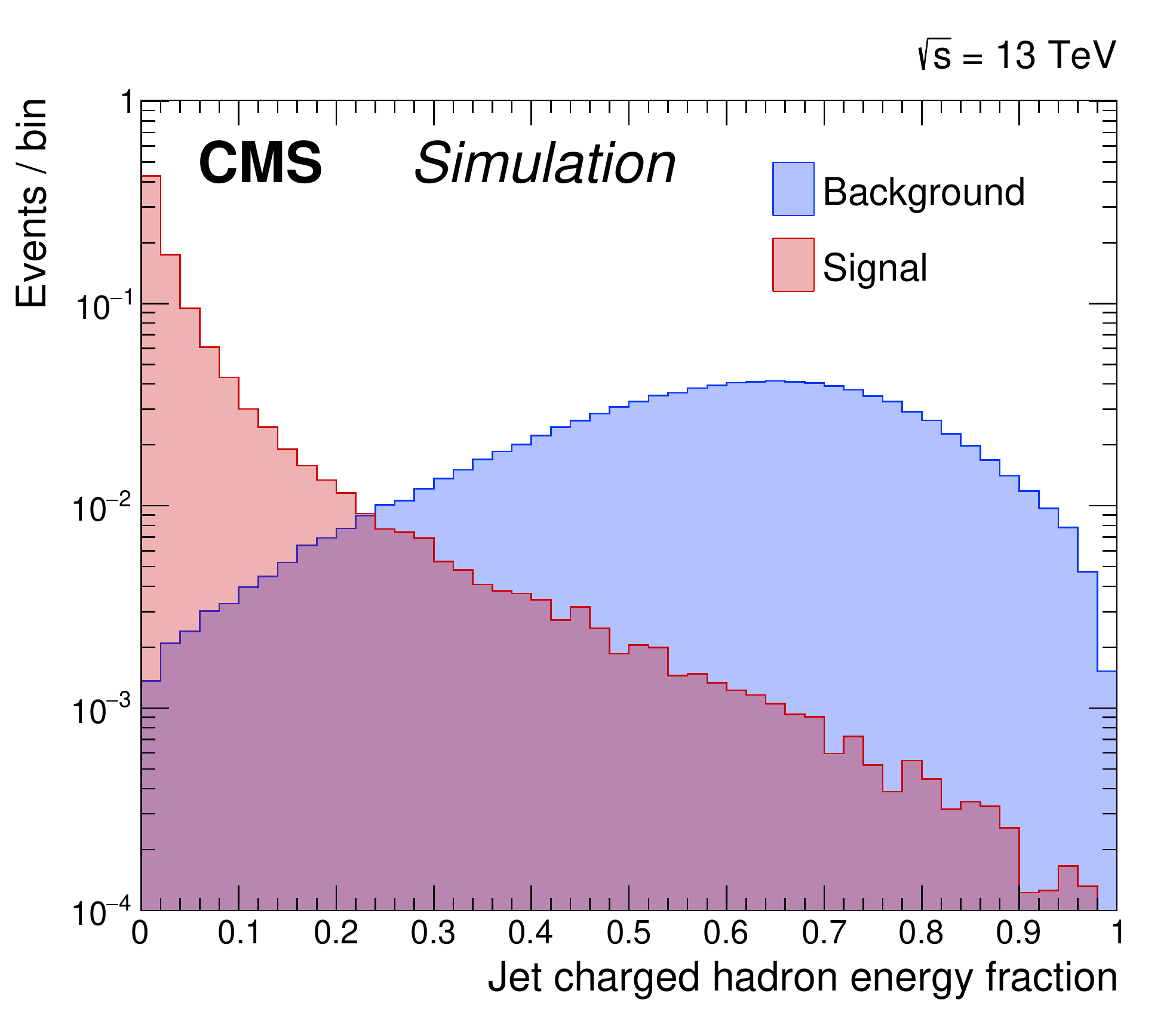}
        \includegraphics[width=0.49\textwidth]{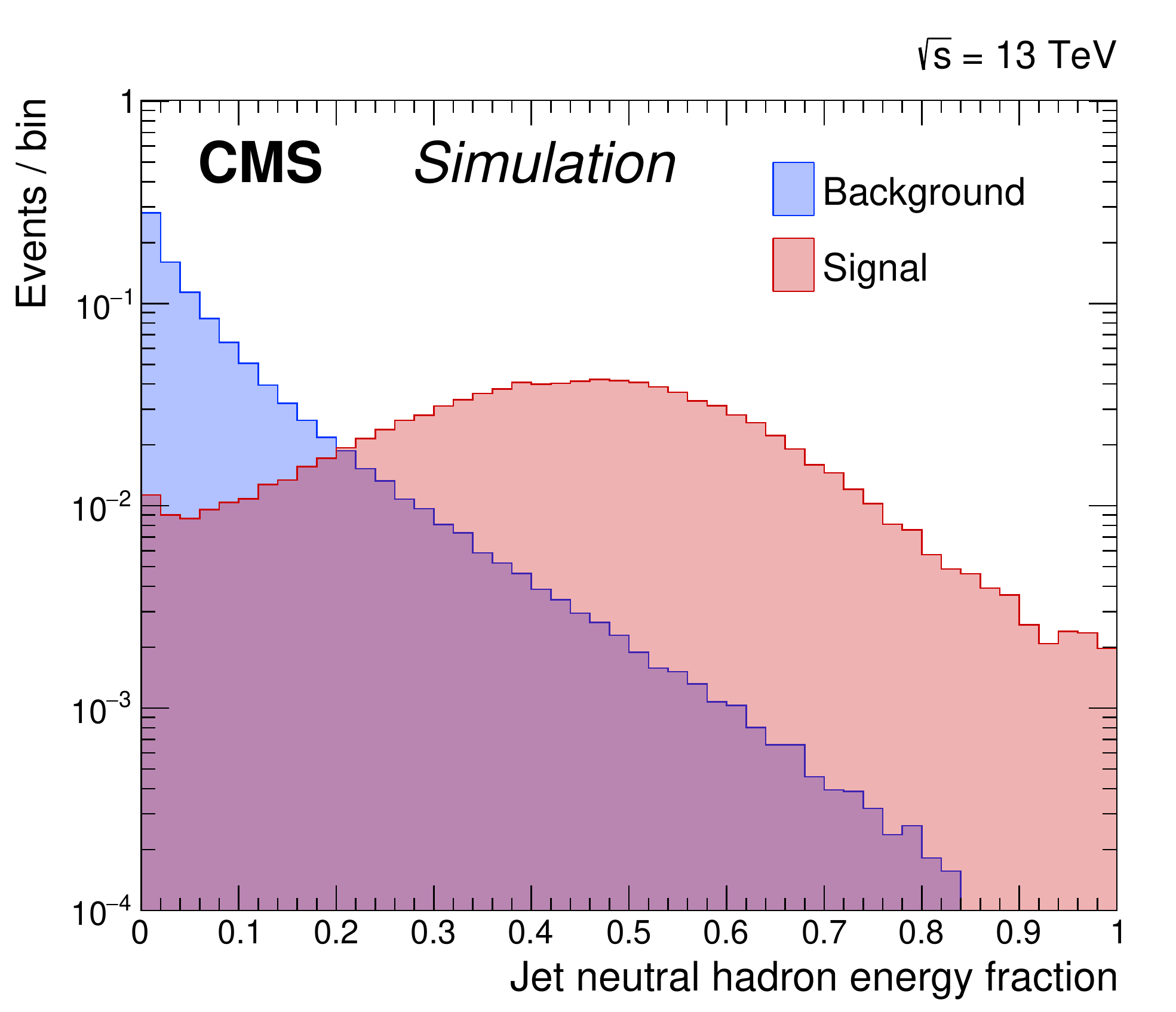}\\
        \includegraphics[width=0.49\textwidth]{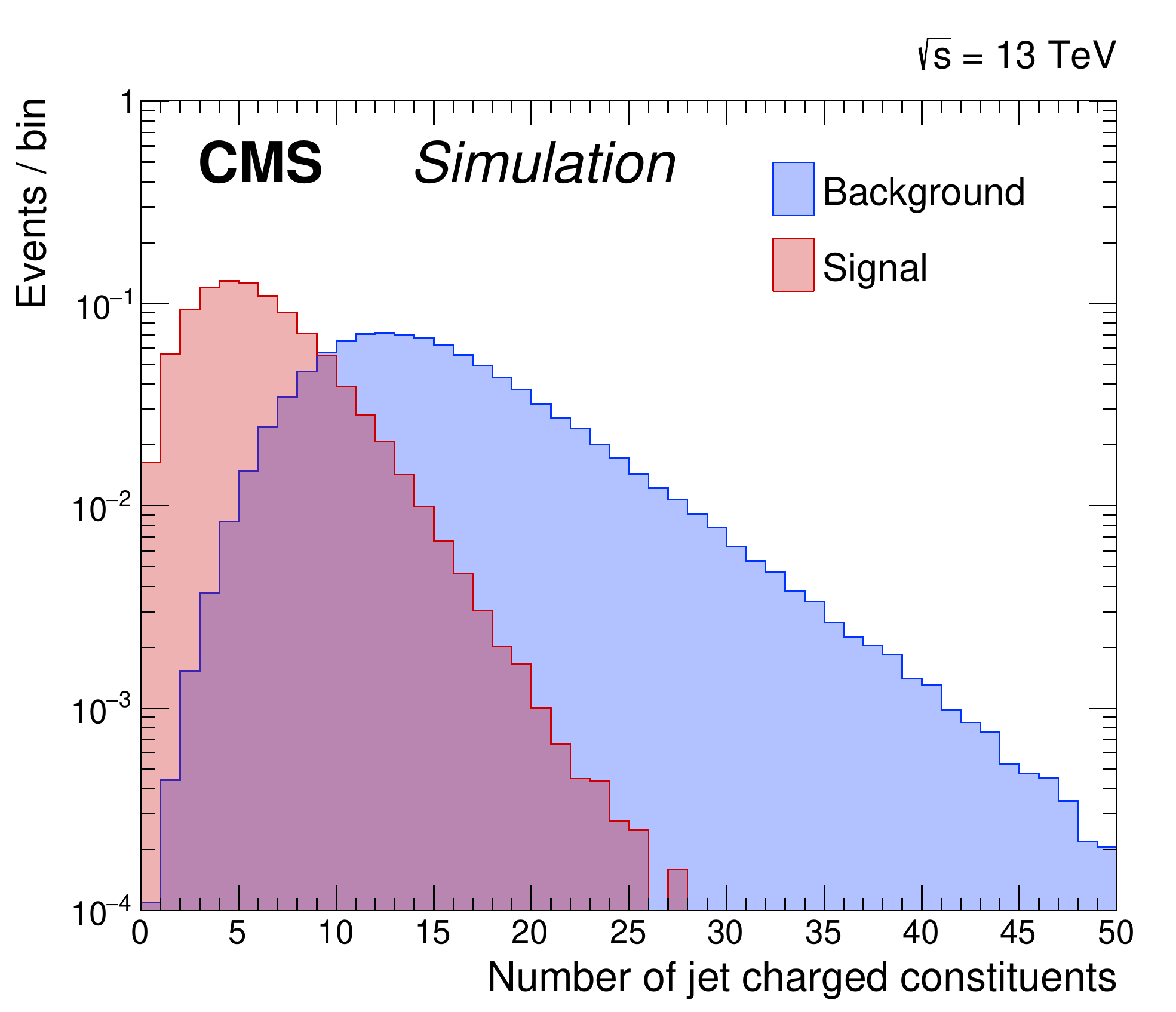}
        \includegraphics[width=0.49\textwidth]{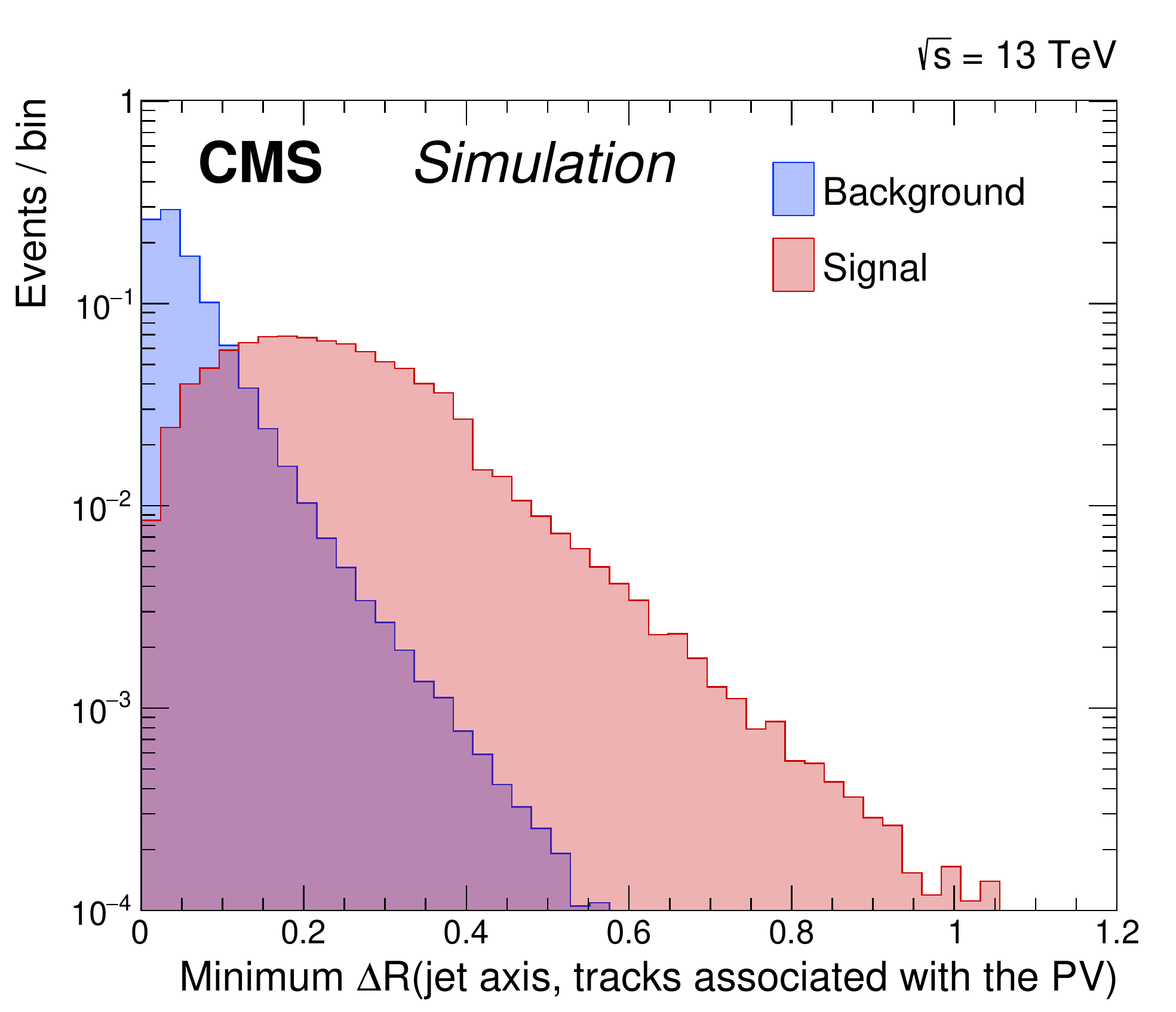}\\
        \includegraphics[width=0.49\textwidth]{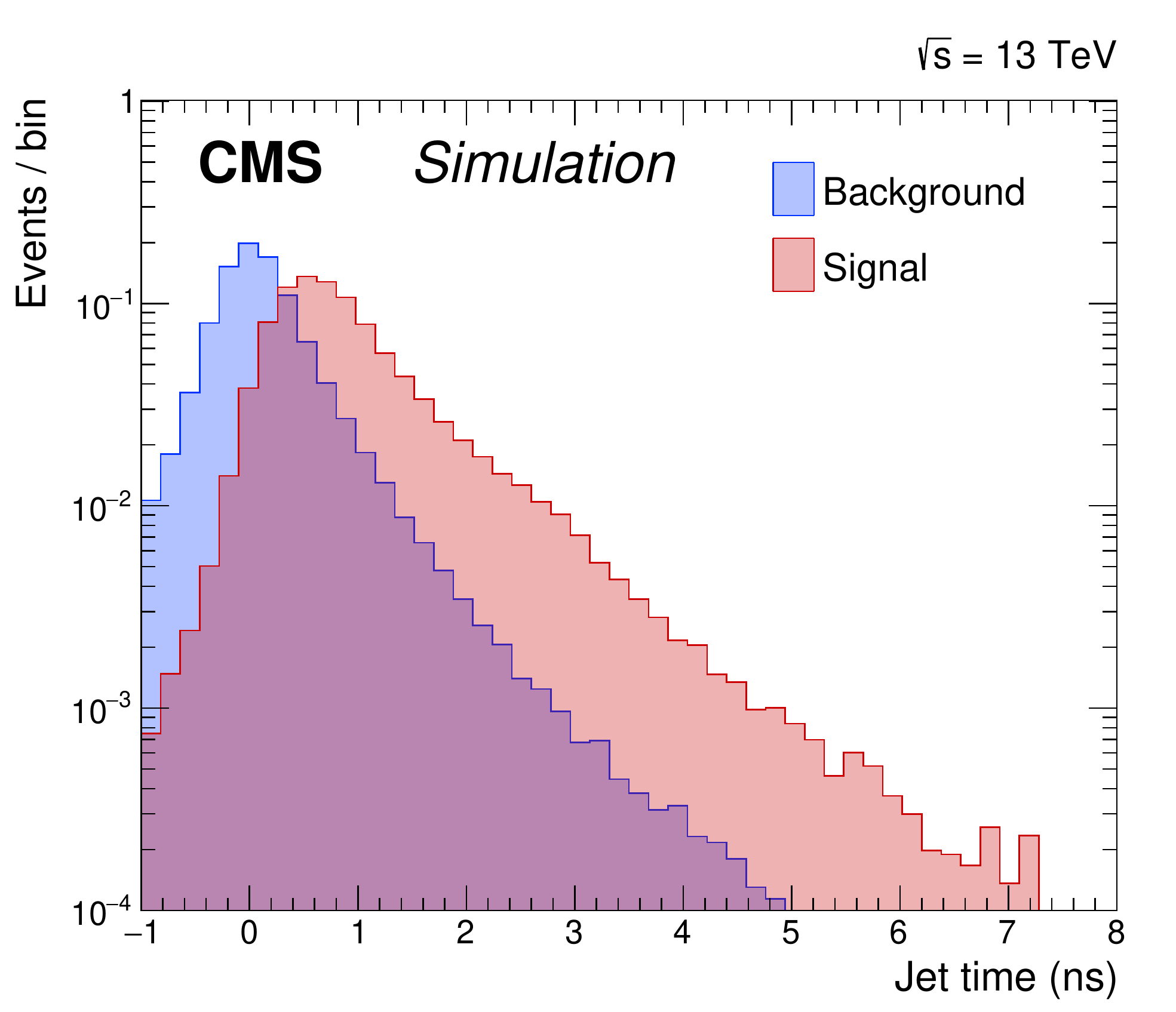}
        \caption{
                The distributions of the most impactful input variables to the TD jet tagger for signal (red, lighter) and collision background (blue, darker).
                They include the charged (upper left) and neutral (upper right) hadron energy fractions, the number of track constituents in the jet (middle left), the $\Delta R$ between the jet axis and the closest track associated with the PV (middle right), and the jet time (lower).
                \label{fig:TaggerInputs}
        }
\end{figure}

\begin{figure}[htpb]
        \centering
        \includegraphics[width=0.49\textwidth]{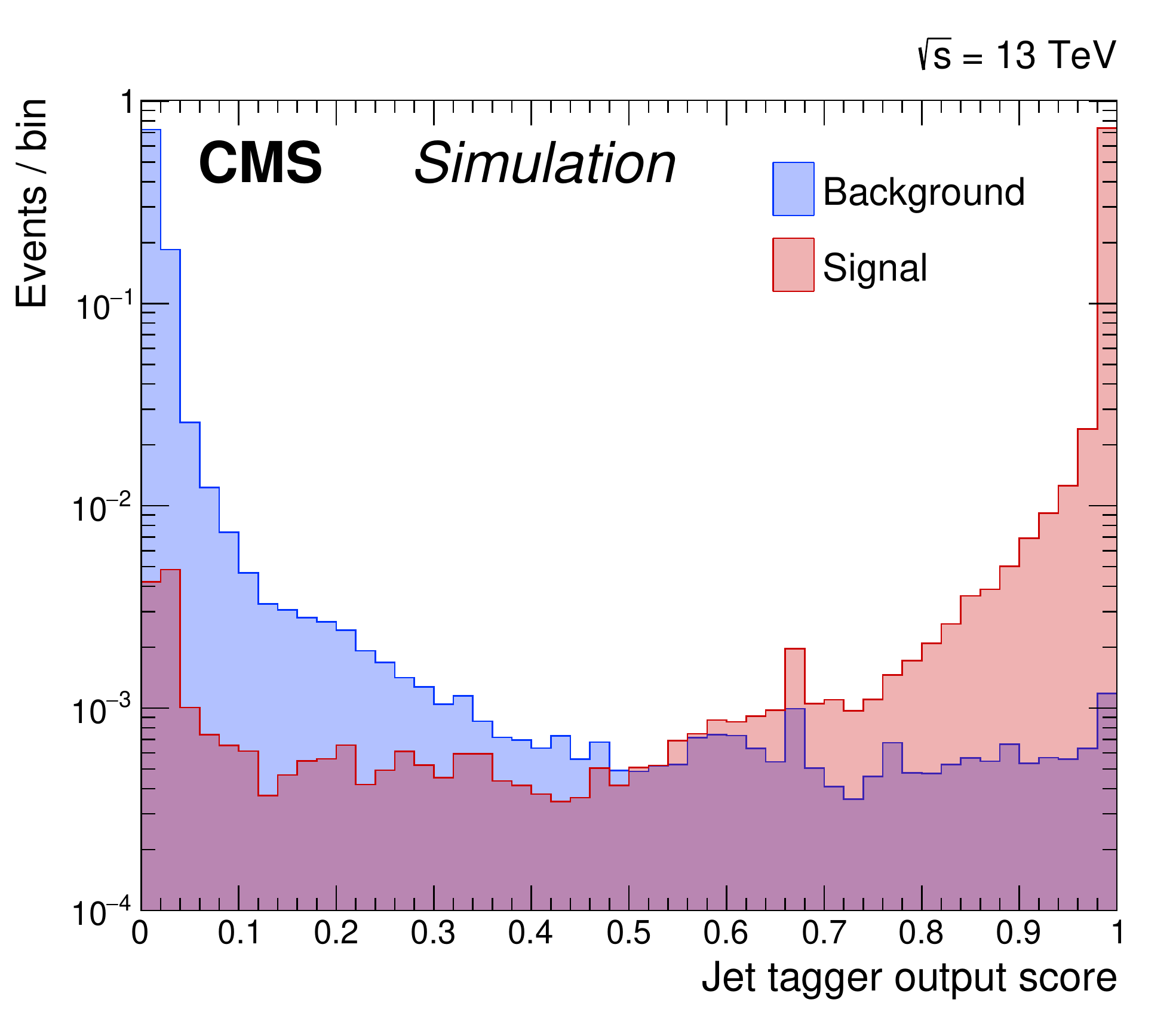}
        \includegraphics[width=0.49\textwidth]{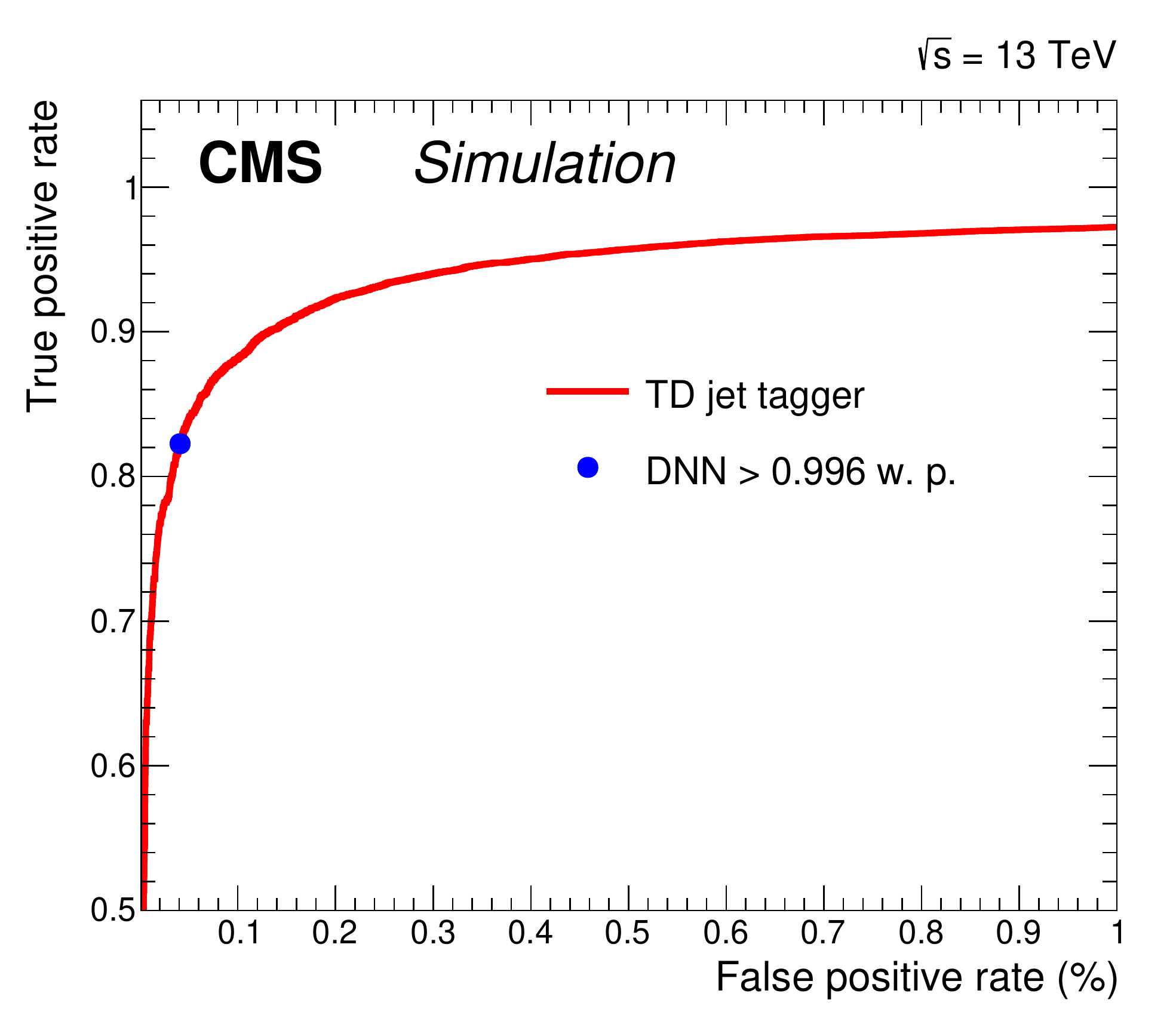}
        \caption{
                TD jet tagger score distributions (left) for signal (red, lighter) and collision background (blue, darker).
                Identification probability for the signal versus the misidentification probability for the background (right) with the tagger working point (w.~p.) used in the analysis shown as a blue marker.
                Both are evaluated using an independent sample of testing events.
                \label{fig:TaggerROC}
        }
\end{figure}

\begin{figure}[htpb]
        \centering
        \includegraphics[width=0.49\textwidth]{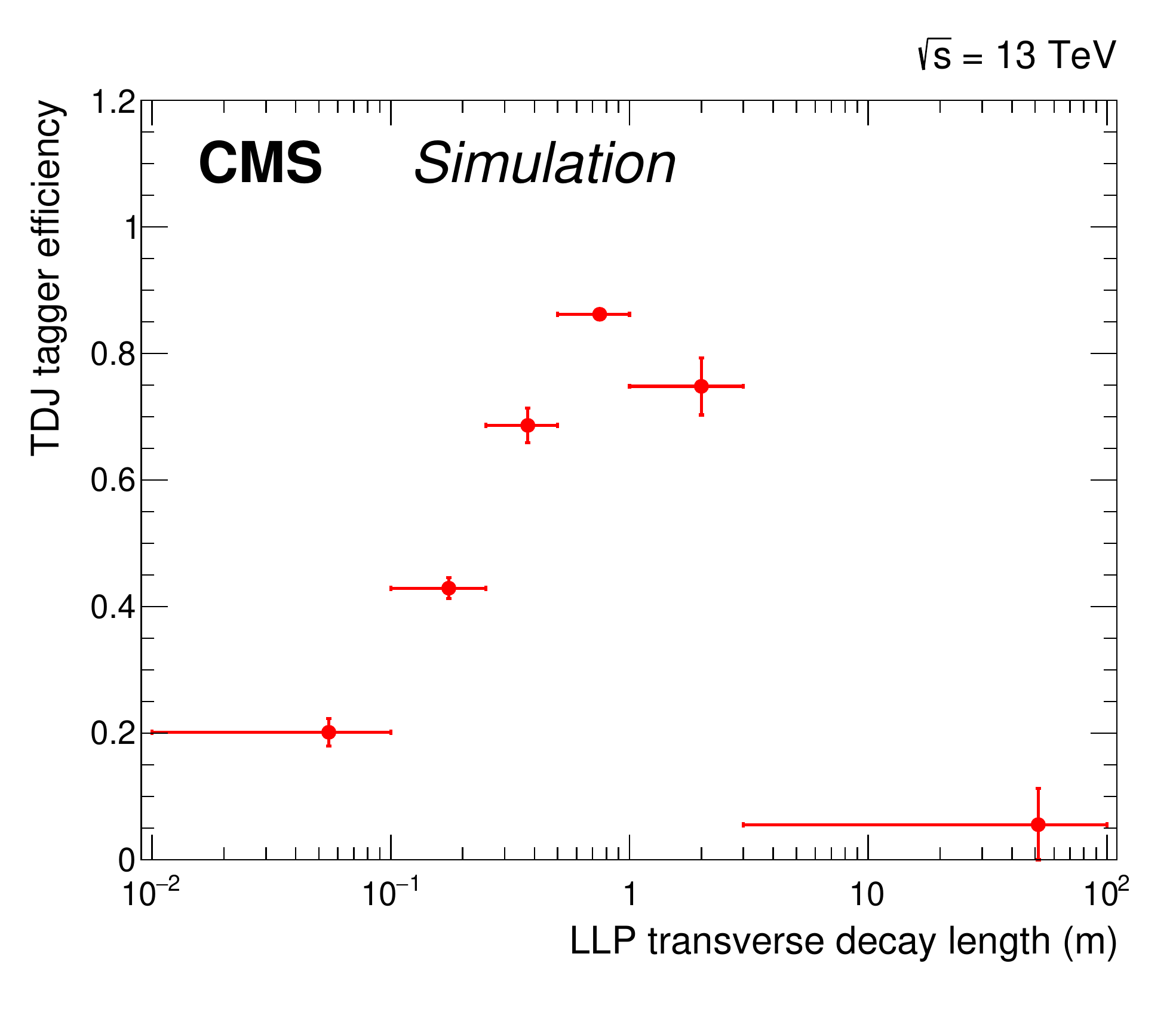}
        \caption{
                The efficiency of the TD jet tagger working point used in the analysis is shown as a function of the lab frame transverse decay length for simulated signals with \PSGczDo mass of 400\GeV. 
                The uncertainties shown account for lifetime dependence and statistical uncertainty.
                \label{fig:TDJTaggerEfficiencyVsDecayLength}
        }
\end{figure}

The most impactful input variables are the charged and neutral hadron energy fractions, the number of track constituents in the jet, the $\Delta R$ between the jet axis and the closest track associated with the PV, and the jet time.
Their distributions for signal and collision background events are shown in Fig.~\ref{fig:TaggerInputs}.
Noncollision backgrounds are not included.
The distribution of the TD jet tagger score for the testing set is shown in Fig.~\ref{fig:TaggerROC} (left), demonstrating the excellent discrimination of the tagger.
The identification probability for the signal versus the misidentification probability for the background is shown in Fig.~\ref{fig:TaggerROC} (right).
In the search, jets are tagged as signal if the TD jet tagger score is larger than 0.996, chosen to yield the best signal sensitivity. 
The signal efficiency of the TD jet tagger is shown in Fig.~\ref{fig:TDJTaggerEfficiencyVsDecayLength} as a function of the transverse decay length in the lab frame, and on average the background efficiency is about $4 \times 10^{-4}$ per candidate jet.

\subsection{Signal efficiency modeling}
\label{sec:taggerEfficiencyModeling}

{\tolerance=800 
The input features of the TD jet tagger making the largest contributions to the signal-to-background discrimination are those related to the trackless nature of signal jets.
For these features, the simulation has not yet been extensively validated. 
To perform this validation and measure the accuracy of the simulation for the TD jet tagger signal efficiency, we use either photons or electrons as proxy objects to represent the LLP decays that occur in the outer tracking or calorimeter volume.
Photons, or electrons with their own tracks removed from consideration, are the best approximations to trackless jets that we can effectively isolate in collision events. 
\par}

Having corrected and validated the simulation of the jet time response as described in Section~\ref{sec:jettime}, we proceed to validate the accuracy of the simulation prediction of the TD tagger discriminator, after factorizing the impact of the jet time measurement.
This is achieved by replicating the time of arrival distribution for signal jets in the corresponding distribution for prompt proxy objects through the introduction of an artificial delay of 1--2\unit{ns}, which is sufficient to cover the bulk of the expected TD tagger scores for signal jets.
Whether to use photons or electrons is based on the jet \pt.
A sample of photons is obtained by selecting $\PZ\to\Plp\Plm\PGg$ events in which one of the leptons ($\Pl = \text{muon or electron}$) radiates a photon.
We require the mass of the $\Plp\Plm\PGg$ system to be between 70--110\GeV to ensure that the sample is not contaminated by jets misidentified as photons.
Because the photon in the $\PZ\to\Plp\Plm\PGg$ sample has a relatively soft momentum spectrum, this control region is only valid for LLP candidate jets with $\pt < 70\GeV$.

\begin{figure}[htpb]
        \centering
        \includegraphics[width=0.49\textwidth]{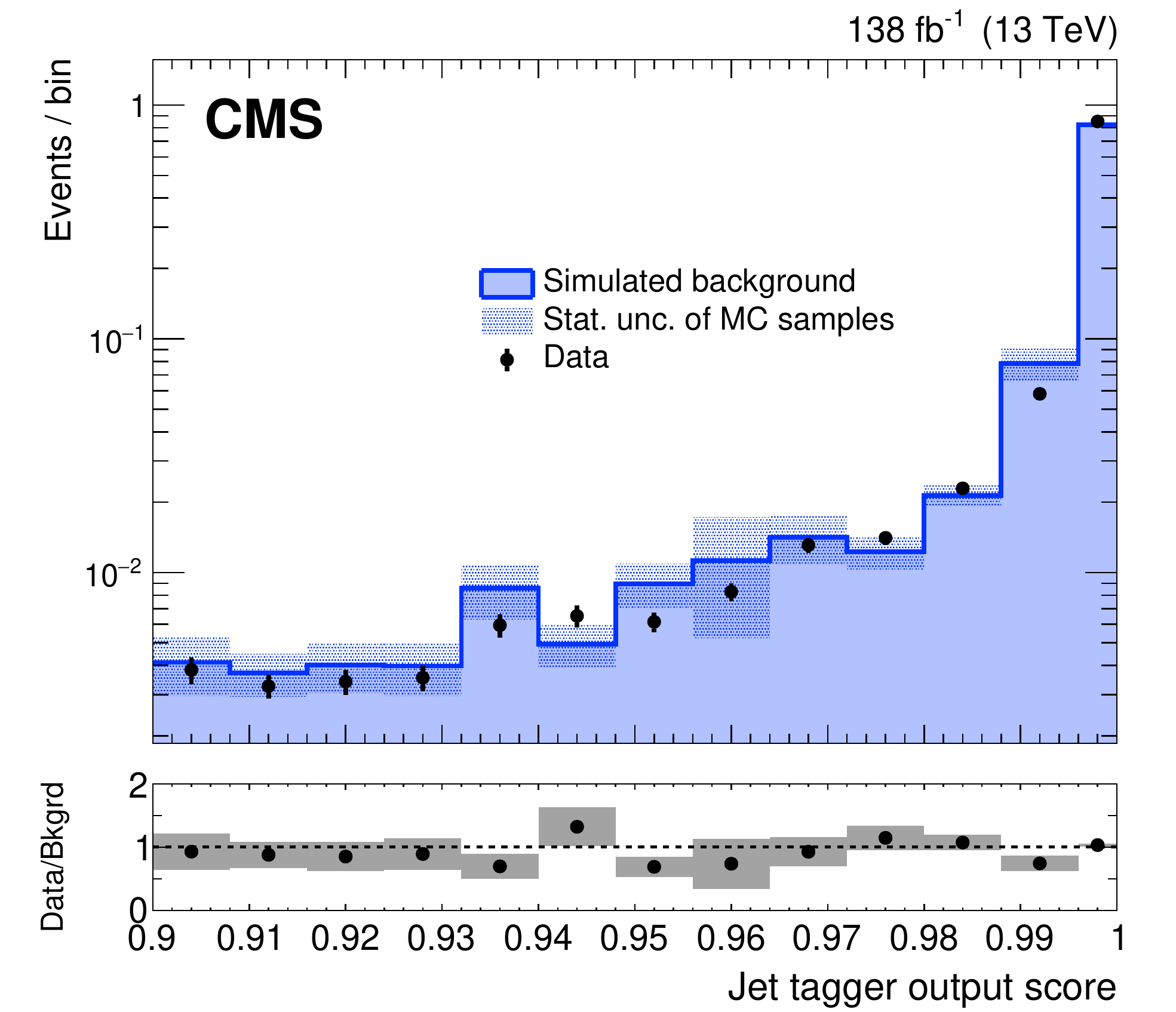}
        \caption{
                The TD jet tagger score distributions for simulation (shaded histogram) and data (black markers) when using electrons from $\PW\to\Pe\Pgne$ events as proxy objects for signal jets. 
                The histograms and data points have been normalized to unit area. 
                The last bin contains jets with tagger scores greater than 0.996, the threshold used to tag signal jets.
                Similar levels of agreement are observed for photon proxy objects from the $\PZ\to\Plp\Plm\PGg$ sample.
                \label{fig:TaggerScore}
        }
\end{figure}

For LLP candidate jets with $\pt > 70\GeV$, we use electrons from $\PW\to\Pe\Pgne$ events as proxy objects obtained by selecting events with an electron passing standard isolation and identification criteria~\cite{CMS:2020uim} and $\ptmiss > 70\GeV$.
The distributions of the TD jet tagger input variables for the photons and electrons in data and simulation are smeared to match the distributions for displaced jets from signal events.
To calculate all the tagger input variables, the track produced by the electron candidate is removed, and the electron is treated as a photon.
We evaluate the TD jet tagger scores for these proxy objects after the corrections to the input variables and imposing the artificial time delay, shown in Fig.~\ref{fig:TaggerScore}, and measure the efficiency for such proxy objects to have a TD jet tagger score larger than 0.996, the threshold used to tag signal jets.
The measured efficiencies are compared to the efficiencies predicted by the simulation, and data-to-simulation correction factors are obtained; these factors show no dependence on jet \pt and $\eta$, and are all in the range 0.9--1.1.
The difference in the correction factor obtained from imposing a delay of 1~and~2\unit{ns} is propagated as a systematic uncertainty.

\section{Background estimation}
\label{sec:bkg}

\begin{figure}[htpb]
        \centering
        \includegraphics[width=0.49\textwidth]{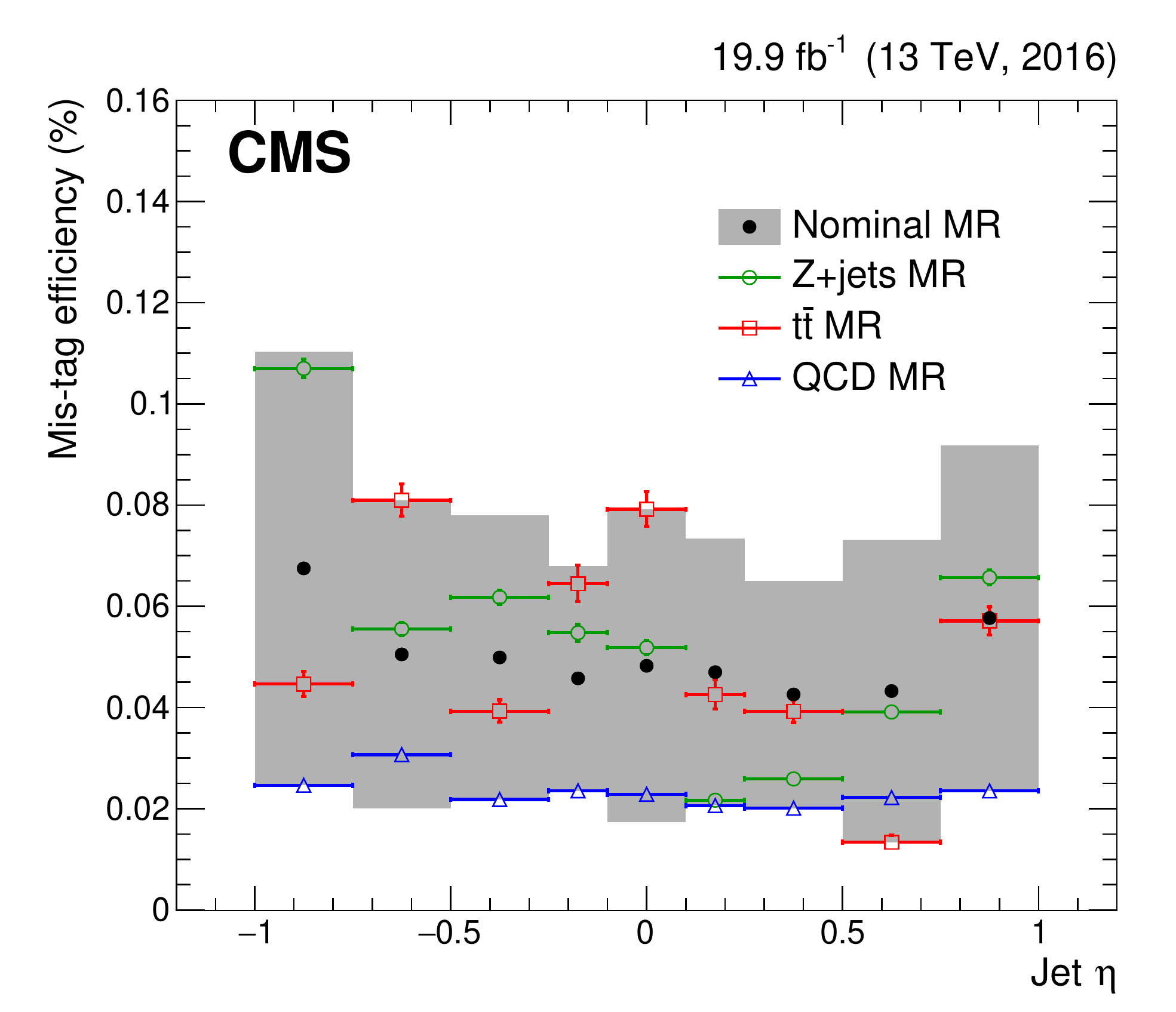}
        \includegraphics[width=0.49\textwidth]{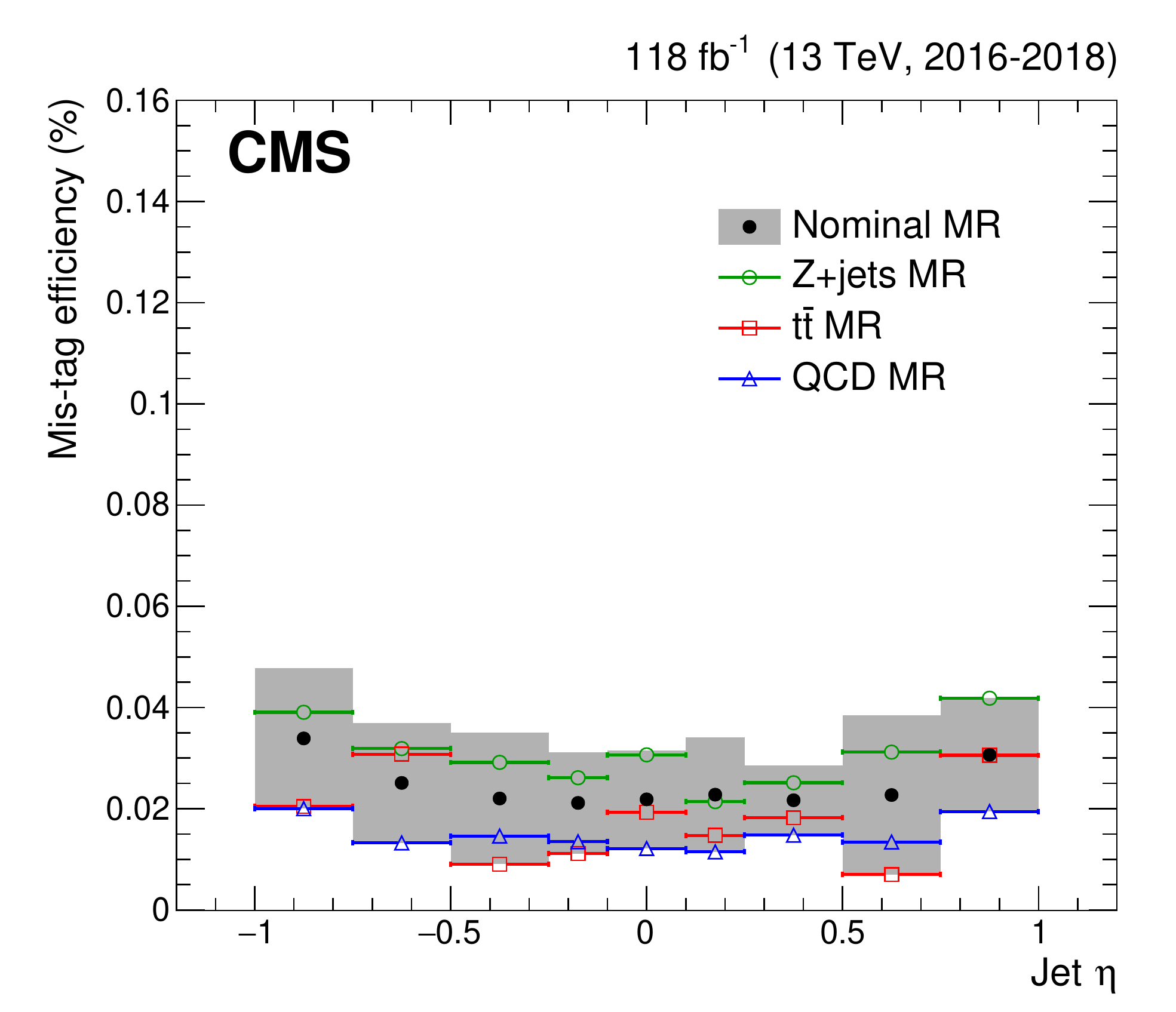}
        \caption{
                The TD jet tagger misidentification probability measured using the nominal \wjets MR (black round markers) is shown along with the systematic uncertainty (gray band), quantifying the degree of process dependence measured from alternative MRs. 
                The measurements in the alternative MRs are displayed as well (\zjets MR as green round markers, \ttbar MR as red squared markers, QCD MR as blue triangular markers) along with their respective statistical uncertainty.
                On the left, this probability is shown for the first 19.9\fbinv of data collected in 2016, while on the right it is shown for the last 16.4\fbinv of data collected in 2016 combined with data collected in 2017--2018.
                \label{fig:MistagEfficiency}
        }
\end{figure}

The main background processes are QCD multijet, \wjets, $\PZ(\PGn\PAGn)$+jets, and \ttbar production, where prompt jets have been misidentified as signal jets by the TD jet tagger.
The misidentification occurs very rarely, less than 0.1\% as shown in Fig.~\ref{fig:MistagEfficiency}, and results primarily from outliers in the jet composition and in the time measurement.
The prompt jets that pass the TD jet tagger either result from tracking inefficiencies or from instances where the fragmentation and hadronization is mostly into photons or neutral hadrons, rendering such jets approximately trackless.
Misidentified jets can also have larger measured times, making them appear to be delayed.
Because the misidentification primarily results from instrumental and resolution effects, the misidentification probability per jet \ebkg does not depend strongly on the process or the presence of other objects in the event.
Motivated by these observations, we employ a matrix method~\cite{ATLAS:2010zaw,CMS:2012exf} to estimate the background contribution to the two-tag SR.

Jets with $\pt > 30\GeV$ are defined as the basic objects used for the matrix method.
We measure \ebkg in signal-depleted measurement regions (MRs) as a function of jet $\eta$.
It is sufficient to parameterize \ebkg only as a function of jet $\eta$ because we observe that it is independent of the jet \pt.
The nominal MR is obtained by selecting events with one electron or muon with $\pt>35$ or 30\GeV, respectively, passing standard isolation and identification criteria~\cite{CMS:2020uim,Sirunyan:2019yvv}.
The events are also required to have $\ptmiss>40\GeV$ to suppress QCD multijet background, and to have transverse mass smaller than 100\GeV to suppress a potential beyond the SM signal.
The \ebkg measurement, derived per jet, is shown in Fig.~\ref{fig:MistagEfficiency}, for two data-taking years.
The misidentification probability is significantly greater for the first 19.9\fbinv of data collected in 2016 because of the impact of suboptimal settings on the tracker readout chip in the presence of large pileup.
The readout chip parameters were reoptimized for the later part of 2016, resulting in improved tracking efficiency and less misidentification for the last 16.4\fbinv of data collected in 2016.

\begin{table}[htb]
        \centering
        \topcaption{
                Definitions of the measurement regions used to quantify the process dependence of the TD jet misidentification probability.
                \label{tab:MRs}
        }
        \begin{tabular}{l|c}
                Measurement region            & Selection                                       \\
                \hline
                \multirow{3}{*}{Nominal}      & 1 electron (muon) with $\pt>35$ (30)\GeV,       \\
                                              & $\ptmiss>40\GeV$,                               \\
                                              & $\mT < 100\GeV$                                 \\[\cmsTabSkip]
                \multirow{3}{*}{\zjets}       & 2 electrons (2 muons) with $\pt>35$ (30)\GeV,   \\
                                              & $60 < m_{\Pe\Pe}$ ($m_{\Pgm\Pgm}$) $< 120\GeV$, \\
                                              & $\ptmiss < 30\GeV$                              \\[\cmsTabSkip]
                \multirow{2}{*}{\ttbar}       & 1 electron and 1 muon with $\pt>35$ and 30\GeV, \\
                                              & $\ptmiss > 30\GeV$                              \\[\cmsTabSkip]
                \multirow{2}{*}{QCD multijet} & 1 jet with $\pt>140\GeV$ online,                \\
                                              & $\ptmiss<25\GeV$ offline                        \\
        \end{tabular}
\end{table}

To quantify the degree of process dependence, misidentification probabilities from three alternative MRs are measured.
The \zjets MR, intended to represent the $\PZ(\PGn\PAGn)$+jets background, is obtained by selecting events with two electrons or two muons, each satisfying the $\pt>35$ or 30\GeV requirement, respectively, and requiring the dilepton mass to be between 60 and 120\GeV.
The \ptmiss is required to be less than 30\GeV to suppress other background processes.
Likewise, the \ttbar MR is obtained by selecting events with one electron and one muon, with the same \pt requirements described above, and $\ptmiss>30\GeV$.
Events in the QCD multijet MR are required to have a jet with $\pt>140\GeV$ online, and $\ptmiss<25\GeV$ offline to avoid signal contamination.
A summary of the different MRs can be found in Table~\ref{tab:MRs}.
The envelope covering the range of misidentification probabilities measured in the nominal and three alternative MRs is used to derive a systematic uncertainty.
This systematic uncertainty quantifying the degree of process dependence is shown in Fig.~\ref{fig:MistagEfficiency} and found to be $45\%$ relative to the collision background and $30\%$ relative to the total background prediction.

To predict the number of background events in the two-tag SR, we consider each untagged jet $i$ with $\pt>30\GeV$ in events passing the preselection with exactly one TD-tagged jet and assign it a probability $\ebkg(\eta_{\text{jet}~i})$ to be misidentified as a TD jet. 
Each untagged jet in the event is treated independently, and the total background prediction \Ntwotagbkg is obtained from the sum
\begin{equation}
        \Ntwotagbkg = \sum_{k=1}^{\Nonetag} \left( \sum^{\Nkuntagged}_{i=1} \ebkg(\eta_{\text{jet}~i}) \right),
        \label{eqn:bkgPredictionFrom1}
\end{equation}
where \Nonetag is the number of one-tag events in data and \Nkuntagged is the number of untagged jets in event $k$.
This prediction ignores contributions of order $\ebkg^{2}$ and is accurate in the limit where \ebkg is very small, which is the case here as \ebkg is below $10^{-3}$. 
Using the matrix method described above, we predict a collision background yield of $0.15\pm0.08\statsyst$ for the two-tag SR.

An alternative background prediction $\Ntwotagbkg^{\prime}$ is obtained by selecting events with zero TD-tagged jets and considering each pair of untagged jets (jets $i$ and $j$) as a potential source of misidentified jets. 
Each untagged jet pair is assigned a weight $\ebkg(\eta_{\text{jet}~i})\ebkg(\eta_{\text{jet}~j})$, and the alternative background prediction is given by
\begin{equation}
        \Ntwotagbkg^{\prime} = \sum_{k=1}^{\Nzerotag} \left( \sum^{\Nkuntagged}_{i=1} \sum^{\Nkuntagged}_{j>i} \ebkg(\eta_{\text{jet}~i})\ebkg(\eta_{\text{jet}~j}) \right),
        \label{eqn:bkgPredictionFrom0}
\end{equation}
where \Nzerotag is the number of zero-tag events in data.

The two background predictions will agree if \ebkg is jet- and process-independent.
Thus, we use the prediction \Ntwotagbkg as the nominal background prediction, and the difference between the two predictions as an estimate of the systematic uncertainty of the method resulting from any additional process dependence.
This systematic uncertainty is found to be 13\%.

Events satisfying the nominal MR selection with two misidentified TD jets are used to define a validation region (VR) in order to test the accuracy of the prediction method.
Using the \ebkg measured in the MR and the same matrix method, we predict $1.1 \pm 0.7$ background events in the VR and observe 1 event in data.
Another validation test is performed by relaxing the jet $\eta$ requirement to $\abs{\eta}<1.48$ in order to decrease the statistical uncertainty of the test.
For the alternative test, we estimate $5.1 \pm 3.1$ events in the VR and observe 4 events in data.
Both tests produce excellent agreement between the predicted and observed yields, thus validating the background estimation method.
Finally, a validation of the $\eta$ dependence of \ebkg is shown in Appendix~\ref{TaggedJetEta}, where the $\eta$ distribution of the TD-tagged jets in the one tag bin of the signal region data is observed to agree with the predicted background.

Although the dominant background contribution is from instrumental misidentification, there are small contributions from noncollision sources including cosmic ray muons and beam halo particles.
To estimate these contributions, we first measure the efficiencies of the cosmic ray and beam halo vetoes in dedicated MRs with high cosmic ray muon and beam halo purities, respectively.
The cosmic ray muon MR is composed of events selected by cosmic ray triggers, enabled during periods when the LHC is not colliding proton beams, and containing a muon that traverses the full detector and a jet.
Events in the beam halo MR are selected by an alternative beam halo filter, which identifies beam halo particles via matching segments in the CSC endcap muon detector. 
For the beam halo MR, we also subtract the instrumental misidentified collision background using the matrix method to prevent the resulting bias on the measured beam halo veto efficiency.
Finally, we invert the cosmic ray muon and beam halo vetoes described in Section~\ref{sec:CosmicsAndBeamHalo} to obtain two separate CRs.
We multiply the corresponding CR event yields by a transfer factor accounting for the measured efficiency of the cosmic ray muon or beam halo veto. 
We obtain predictions for the cosmic ray muon and beam halo backgrounds of $0.03 \pm 0.02$ and $0.05 \pm 0.05$ events, respectively, consistent with zero events in each case.

\section{Systematic uncertainties}
\label{sec:systematics}

The source of the dominant background uncertainty is the estimation of the misidentified TD jet background.
This uncertainty includes contributions related to the CR sample size, \ebkg process dependence, and the matrix method validity, amounting to 4, 30, and 13\%, respectively, relative to the total background prediction.
The noncollision background uncertainties correspond to about 23\% relative to the total background prediction.
Several systematic uncertainties in the signal yield prediction are also considered.
The uncertainty in the TD jet tagger efficiency, which includes a component from the jet time correction uncertainty as well as a component estimated using the methods discussed in Section~\ref{sec:taggerEfficiencyModeling}, is the largest and amounts to up to 29\% relative to the signal yield, depending on the signal model parameters.
The integrated luminosities for the 2016, 2017, and 2018 data-taking years have 1.2--2.5\% individual uncertainties~\cite{CMS-LUM-17-003,CMS-PAS-LUM-17-004,CMS-PAS-LUM-18-002}, while the overall uncertainty for the 2016--2018 period is 1.6\%.
The impact of uncertainties related to jet energy scale and resolution~\cite{CMS-PAS-JME-16-003,CMS:2019ctu}, PDFs, missing higher order QCD corrections, pileup modeling, integrated luminosity, simulation sample size, and lepton and photon veto efficiency is also estimated and summarized in Table~\ref{tab:Systematics}.

\begin{table}[htb]
        \centering
        \topcaption{
                Summary of combined statistical and systematic uncertainties, the size of their effect, and whether it applies to the signal or background yield predictions.
                Ranges for signal systematic uncertainties reflect their impact on different signal parameter space points.
                \label{tab:Systematics}
        }
        \begin{tabular}{lcc}
                Uncertainty source                                 & Process    & Uncertainty [\%] \\
                \hline
                Background CR sample size                          & Background & 4                \\
                TD jet tagger misidentification process dependence & Background & 30               \\
                Background estimation method                       & Background & 13               \\[\cmsTabSkip]
                Noncollision background                            & Background & 23               \\
                TD jet tagger efficiency                           & Signal     & 8--29            \\
                Jet energy scale                                   & Signal     & 0.1--11          \\
                Jet energy resolution                              & Signal     & 0.2--10          \\
                PDFs                                               & Signal     & $\le$1           \\
                Missing higher-order QCD corrections               & Signal     & $\le$1           \\
                Pileup                                             & Signal     & 0.3--6.3         \\
                Integrated luminosity                              & Signal     & 2.5              \\
                Signal sample size                                 & Signal     & 5--8             \\
                Lepton and photon veto efficiency                  & Signal     & $\ll$1           \\
        \end{tabular}
\end{table}

\section{Results and interpretation}
\label{sec:results}

\begin{figure}[htb]
        \centering
        \includegraphics[width=0.59\textwidth]{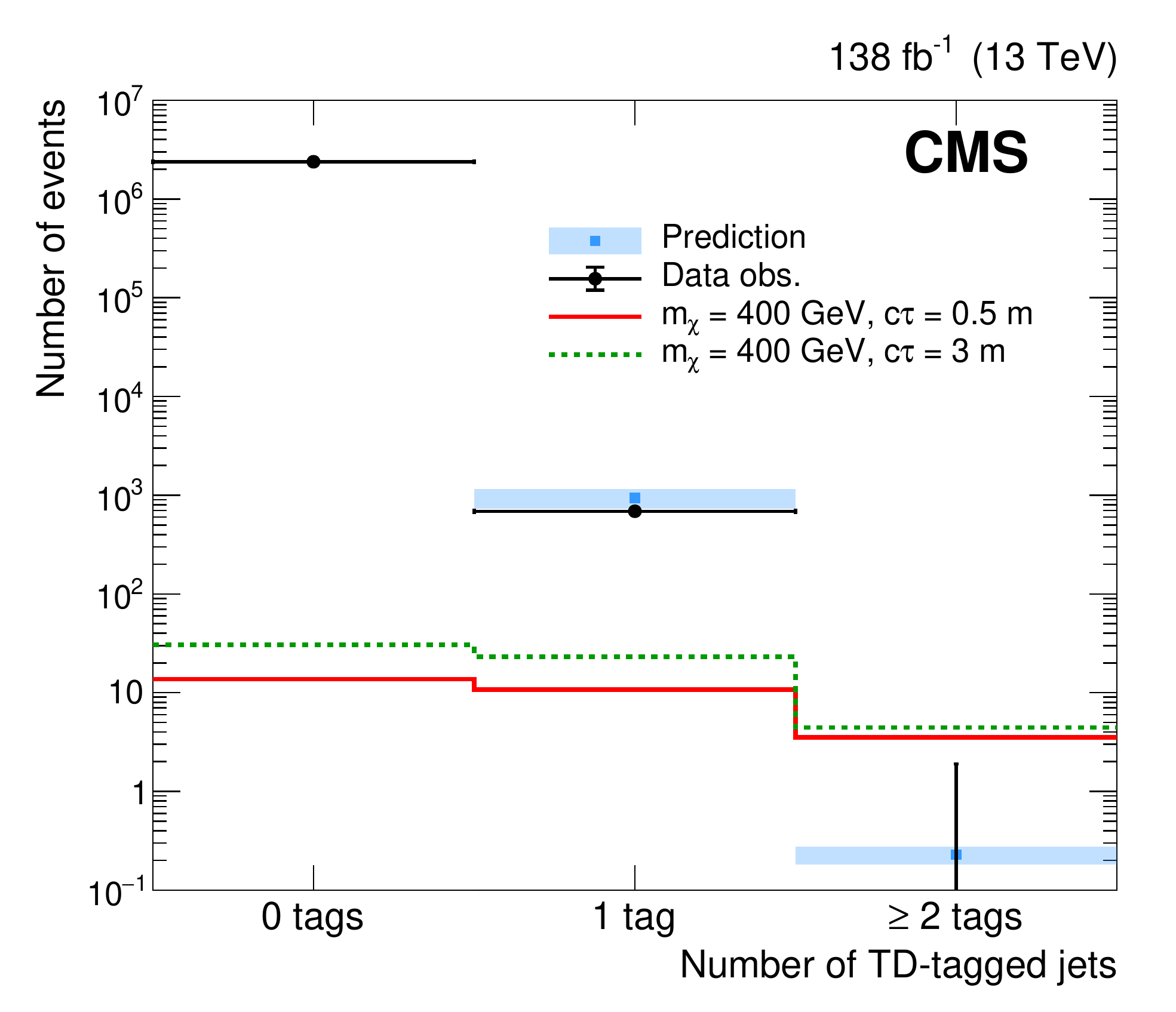}
        \caption{
                Distribution of the number of TD-tagged jets for the $\mChi = 400\GeV$ simulated signal samples with $\ctauChi=0.5\unit{m}$ (solid red line) and $\ctauChi=3.0\unit{m}$ (dotted green line), estimated background (blue square markers), and data (black round markers). The signal distributions are normalized to the expected cross section limit.
                The blue shaded region indicates the systematic uncertainty in the background prediction.
                No background prediction is shown for the bin with zero TD-tagged jets as it is the main control region used to predict the background for the other two bins.
                There are zero observed events in the bin with two or more TD-tagged jets.
                \label{fig:nTDJ}
        }
\end{figure}

Figure~\ref{fig:nTDJ} shows the observed data for the three bins corresponding to $\NT = 0$, 1, and ${\geq}2$, and the estimated background yields for the latter two $\NT$ bins.
In the most sensitive SR bin, corresponding to $\NT \geq 2$, we observe 0 events with an expected background of $0.23\pm0.10$ events.

We interpret the results in the context of a simplified model of GMSB chargino-neutralino production~\cite{Dimopoulos:1996vz,Matchev:1999ft}.
As discussed in Section~\ref{sec:intro}, the long-lived neutralinos are effectively pair produced and decay to either $\PH\PXXSG$ or $\PZ\PXXSG$.
The branching fraction $\cbf(\ChiToHG)$ is varied from 0 to 1, assuming $\cbf(\ChiToHG) + \cbf(\ChiToZG) = 1.0$

We use a test statistic based on the profile likelihood ratio~\cite{ATLAS:2011tau} to extract the signal.
Upper limits at 95\% confidence level (\CL) on the product of the neutralino pair production cross section (\xsecChi) and the relevant branching fraction are derived using the modified frequentist \CLs criterion~\cite{Junk:1999kv,Read:2002hq} and an asymptotic formulae~\cite{Cowan:2010js}.
Systematic uncertainties are incorporated into the analysis via nuisance parameters with log-normal probability density functions and are treated according to the frequentist paradigm.

\begin{figure}[htpb]
        \centering
        \includegraphics[width=0.49\textwidth]{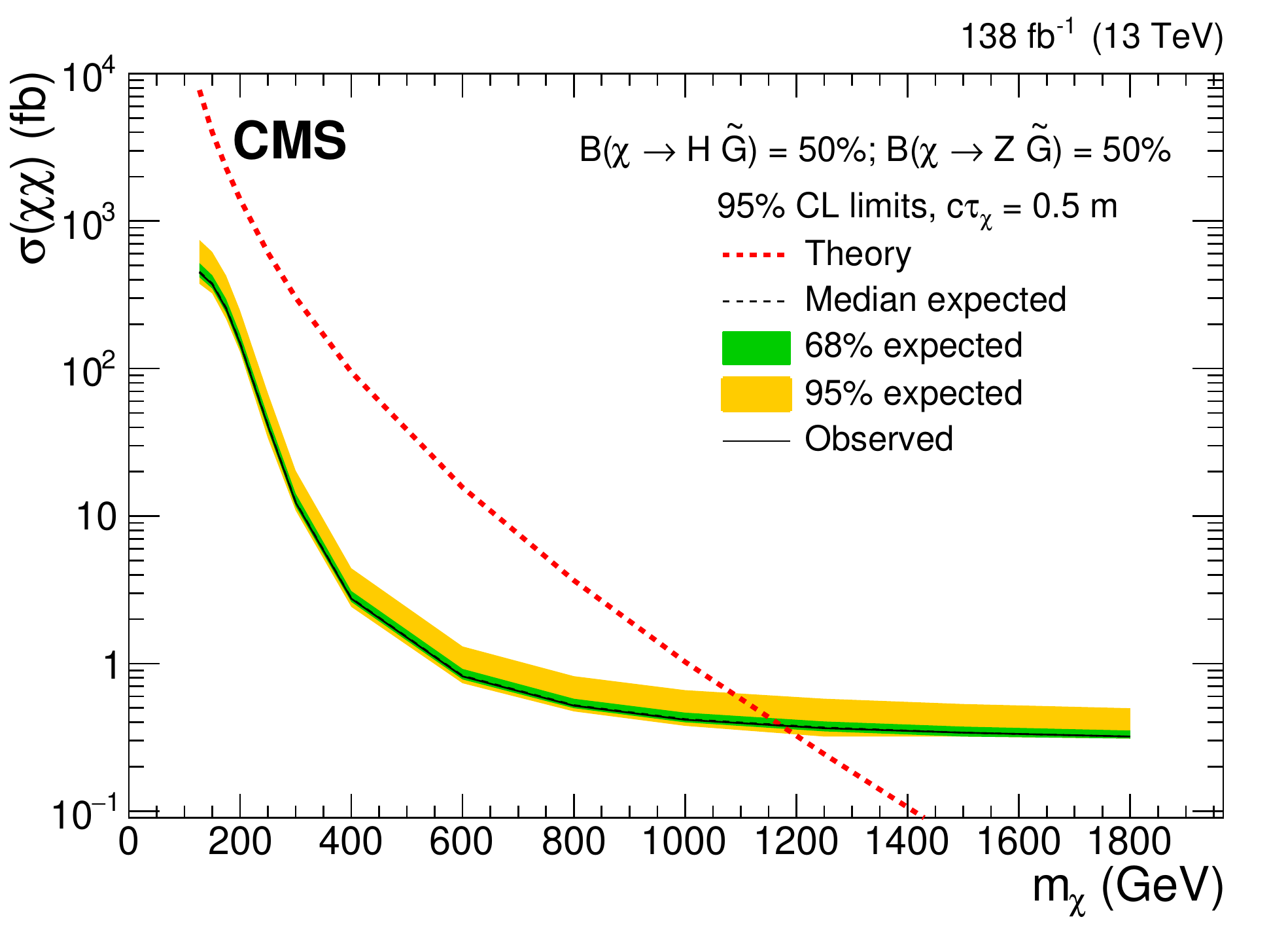}
        \includegraphics[width=0.49\textwidth]{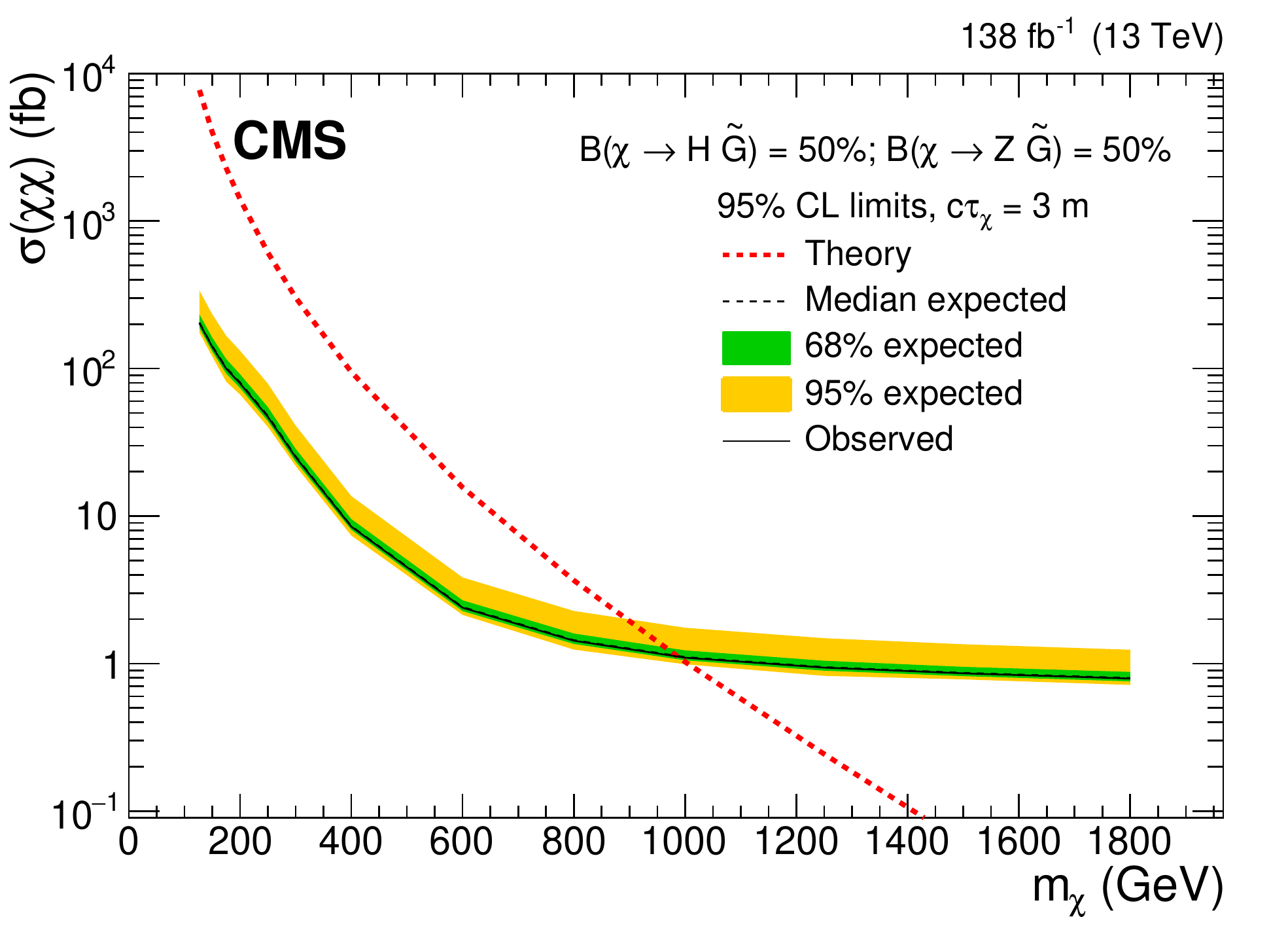}
        \caption{
                Expected and observed 95\% \CL upper limits on \xsecChi as functions of \mChi in a scenario with $\cbf(\ChiToHG) = 0.5$ and $\ctau = 0.5\unit{m}$ (left) or 3\unit{m} (right).
                \label{fig:LimitsVsMass}
        }
\end{figure}

\begin{figure}[htpb]
        \centering
        \includegraphics[width=0.49\textwidth]{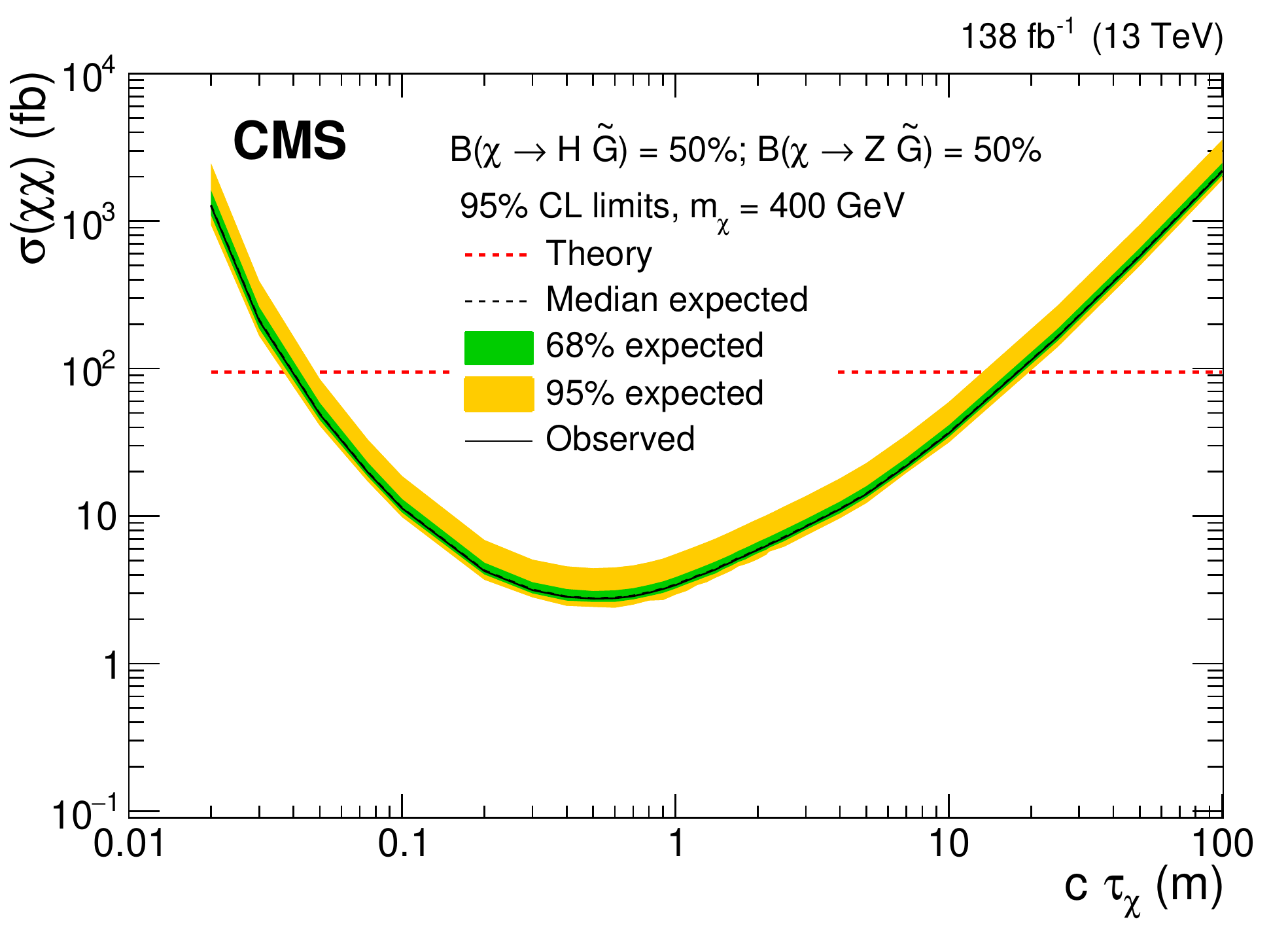}
        \includegraphics[width=0.49\textwidth]{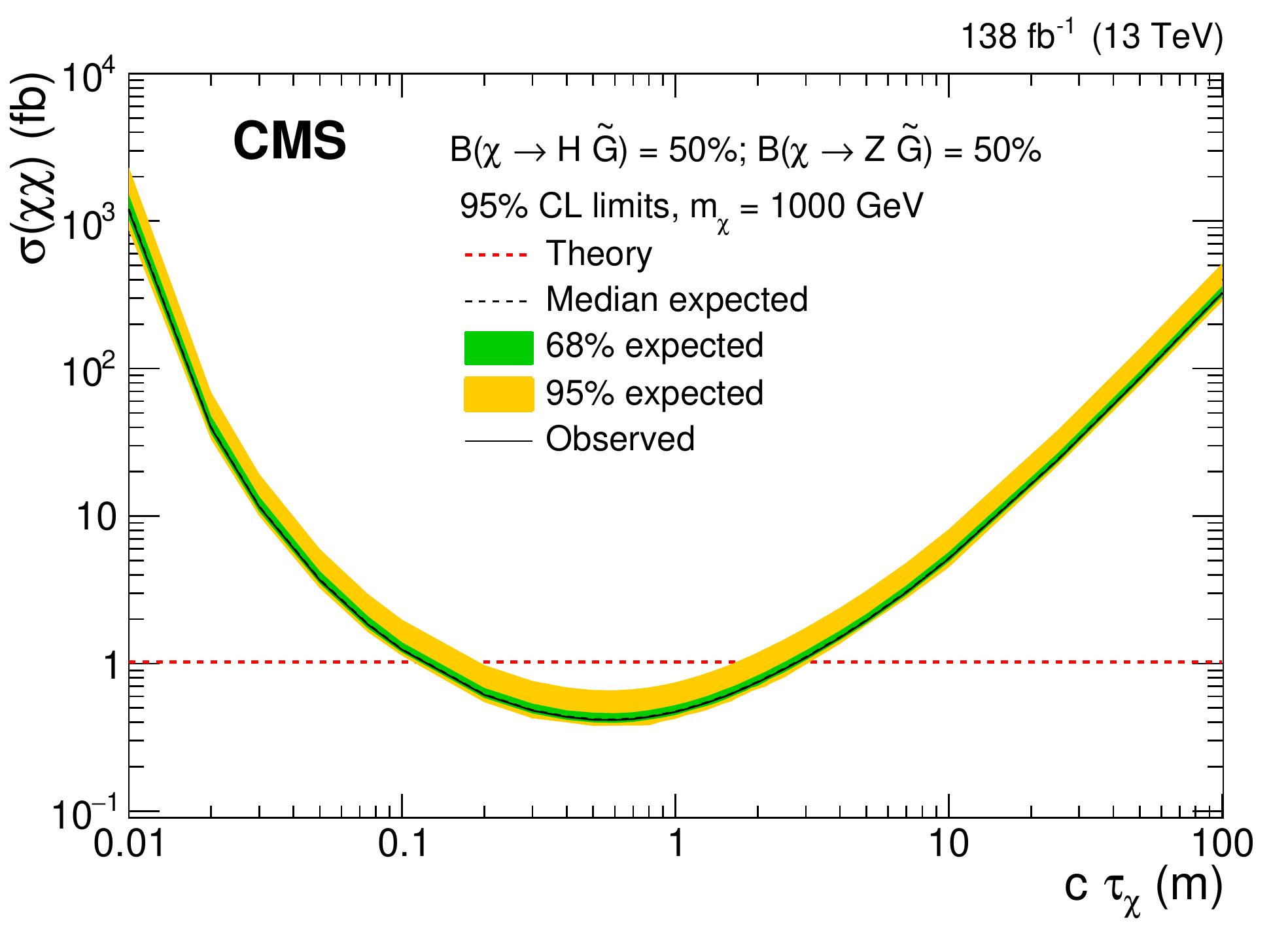}
        \caption{
                Expected and observed 95\% \CL upper limits on \xsecChi as functions of \ctauChi in a scenario with $\cbf(\ChiToHG) = 0.5$ and $\mChi=400\GeV$ (left) or 1000\GeV (right).
                \label{fig:LimitsVsCtau}
        }
\end{figure}

\begin{figure}[htpb]
        \centering
        \includegraphics[width=0.8\textwidth]{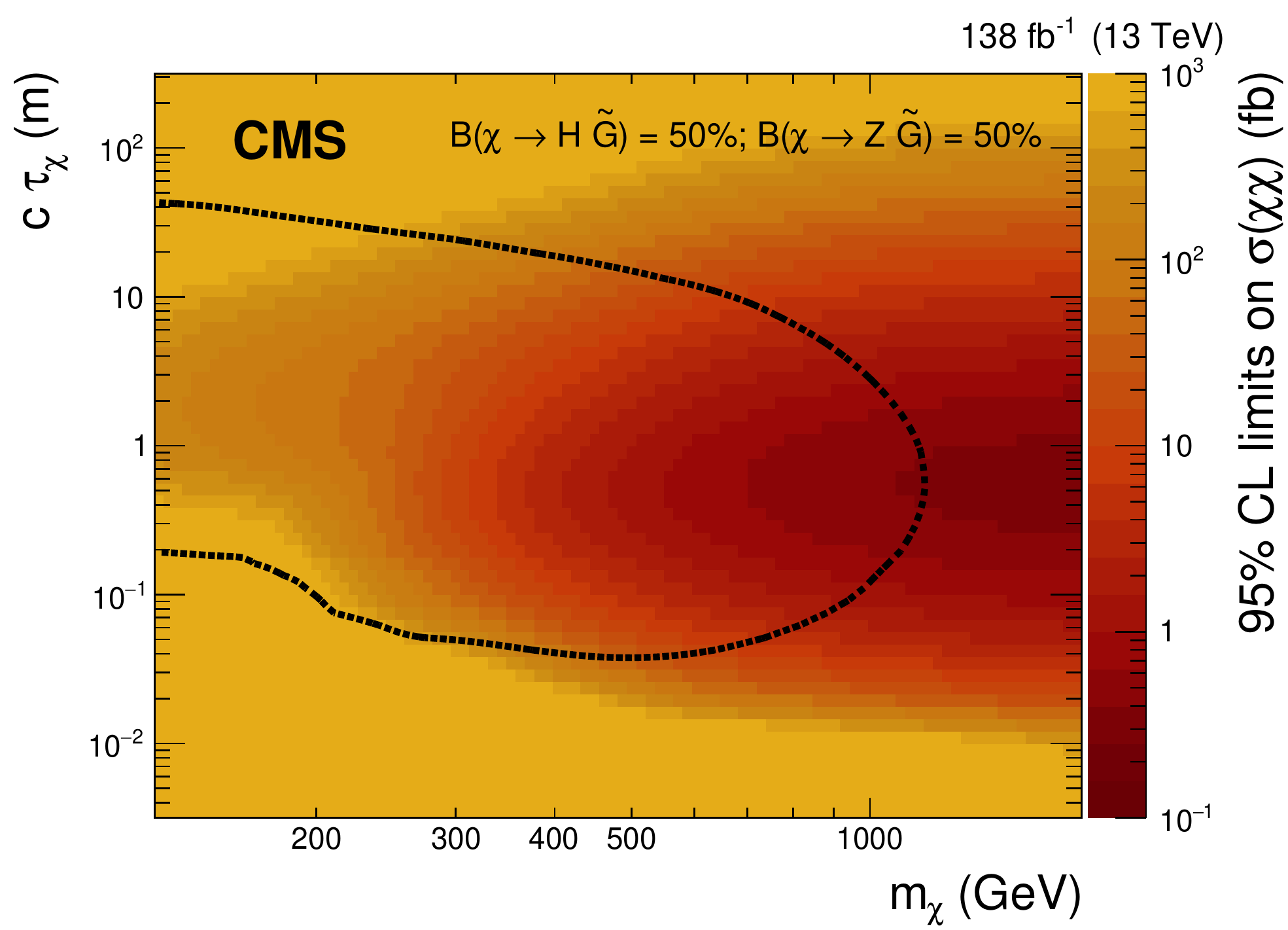}
        \caption{
                The observed 95\% \CL upper limit on \xsecChi as a function of \mChi and \ctauChi in a scenario with $\cbf(\ChiToHG) = 0.5$. 
                The area enclosed by the dotted black line corresponds to the observed excluded region.
                \label{fig:Limits2D}
        }
\end{figure}

Figures~\ref{fig:LimitsVsMass}~and~\ref{fig:LimitsVsCtau} show the upper limits on \xsecChi as functions of the \PSGczDo mass (\mChi) and proper decay length (\ctauChi), respectively, assuming $\cbf(\ChiToHG) = 0.5$.
Figure~\ref{fig:Limits2D} shows the observed upper limits as a function of both \mChi and \ctauChi.
The upper limits are relatively independent of the branching fraction $\cbf(\ChiToHG)$, varying by less than 10\% over the full range.
For example, for $\ctauChi=0.5\unit{m}$ we exclude cross sections of 160, 2.6, and 0.8\unit{fb} for \mChi of 200, 400, and 600\GeV, respectively.
Compared to previous searches for promptly decaying \PSGczDo in the same simplified model~\cite{CMS:2017nin}, the sensitivity of the current search expressed in terms of the cross section limit is about 20 (10) times better at $\mChi = 400$ (600)\GeV.
At a 95\% \CL, we exclude \PSGczDo masses up to approximately 1180 (990)\GeV when \ctauChi is 0.5 (3.0)\unit{m}.
For $\mChi=400$ (1000)\GeV, we exclude \ctauChi in the range from 0.04 to 20 (0.1 to 3)\unit{m}.
Tabulated results are provided in the HEPData record for this analysis~\cite{hepdata}.

\clearpage

\section{Summary}
\label{sec:summary}

A search for long-lived particles has been carried out using proton-proton collision data at $\sqrt{s} = 13\TeV$, corresponding to an integrated luminosity of 138\fbinv, using missing transverse momentum and a novel and highly discriminating deep neural network tagger for trackless and delayed (TD) jets.
Each additional TD-tagged jet required suppresses standard model background processes by more than three orders of magnitude while maintaining the signal efficiency above 80\%.
A background estimation method based on control samples in data uses the tagger's measured misidentification probability to extrapolate from event samples with one or fewer tagged jets to the signal region comprising events with two or more tagged jets.
The results are interpreted in the context of a simplified model of electroweak production of chargino-neutralino pairs.
For a neutralino (\PSGczDo) proper decay length of $\ctauChi=0.5\unit{m}$, we exclude cross sections of 160, 2.6, and 0.8\unit{fb} for \PSGczDo masses (\mChi) of 200, 400, and 600\GeV, respectively, at 95\% confidence level.
Compared to previous searches for promptly decaying \PSGczDo in the same simplified model, the sensitivity of the current search expressed in terms of cross section limit is about 20 (10) times better at $\mChi = 400$ (600)\GeV.
In the case of a long-lived \PSGczDo with $\ctauChi=0.5\unit{m}$, \PSGczDo masses up to 1.18\TeV are excluded at 95\% confidence level.
The current search is the best result to date in the mass range from the kinematic limit imposed by the Higgs boson mass up to 1.8\TeV.       

\begin{acknowledgments}
We congratulate our colleagues in the CERN accelerator departments for the excellent performance of the LHC and thank the technical and administrative staffs at CERN and at other CMS institutes for their contributions to the success of the CMS effort. In addition, we gratefully acknowledge the computing centers and personnel of the Worldwide LHC Computing Grid and other centers for delivering so effectively the computing infrastructure essential to our analyses. Finally, we acknowledge the enduring support for the construction and operation of the LHC, the CMS detector, and the supporting computing infrastructure provided by the following funding agencies: BMBWF and FWF (Austria); FNRS and FWO (Belgium); CNPq, CAPES, FAPERJ, FAPERGS, and FAPESP (Brazil); MES and BNSF (Bulgaria); CERN; CAS, MoST, and NSFC (China); MINCIENCIAS (Colombia); MSES and CSF (Croatia); RIF (Cyprus); SENESCYT (Ecuador); MoER, ERC PUT and ERDF (Estonia); Academy of Finland, MEC, and HIP (Finland); CEA and CNRS/IN2P3 (France); BMBF, DFG, and HGF (Germany); GSRI (Greece); NKFIH (Hungary); DAE and DST (India); IPM (Iran); SFI (Ireland); INFN (Italy); MSIP and NRF (Republic of Korea); MES (Latvia); LAS (Lithuania); MOE and UM (Malaysia); BUAP, CINVESTAV, CONACYT, LNS, SEP, and UASLP-FAI (Mexico); MOS (Montenegro); MBIE (New Zealand); PAEC (Pakistan); MES and NSC (Poland); FCT (Portugal); MESTD (Serbia); MCIN/AEI and PCTI (Spain); MOSTR (Sri Lanka); Swiss Funding Agencies (Switzerland); MST (Taipei); MHESI and NSTDA (Thailand); TUBITAK and TENMAK (Turkey); NASU (Ukraine); STFC (United Kingdom); DOE and NSF (USA).
         
\hyphenation{Rachada-pisek} Individuals have received support from the Marie-Curie program and the European Research Council and Horizon 2020 Grant, contract Nos.\ 675440, 724704, 752730, 758316, 765710, 824093, 884104, and COST Action CA16108 (European Union); the Leventis Foundation; the Alfred P.\ Sloan Foundation; the Alexander von Humboldt Foundation; the Belgian Federal Science Policy Office; the Fonds pour la Formation \`a la Recherche dans l'Industrie et dans l'Agriculture (FRIA-Belgium); the Agentschap voor Innovatie door Wetenschap en Technologie (IWT-Belgium); the F.R.S.-FNRS and FWO (Belgium) under the ``Excellence of Science -- EOS" -- be.h project n.\ 30820817; the Beijing Municipal Science \& Technology Commission, No. Z191100007219010; the Ministry of Education, Youth and Sports (MEYS) of the Czech Republic; the Hellenic Foundation for Research and Innovation (HFRI), Project Number 2288 (Greece); the Deutsche Forschungsgemeinschaft (DFG), under Germany's Excellence Strategy -- EXC 2121 ``Quantum Universe" -- 390833306, and under project number 400140256 - GRK2497; the Hungarian Academy of Sciences, the New National Excellence Program - \'UNKP, the NKFIH research grants K 124845, K 124850, K 128713, K 128786, K 129058, K 131991, K 133046, K 138136, K 143460, K 143477, 2020-2.2.1-ED-2021-00181, and TKP2021-NKTA-64 (Hungary); the Council of Science and Industrial Research, India; the Latvian Council of Science; the Ministry of Education and Science, project no. 2022/WK/14, and the National Science Center, contracts Opus 2021/41/B/ST2/01369 and 2021/43/B/ST2/01552 (Poland); the Funda\c{c}\~ao para a Ci\^encia e a Tecnologia, grant CEECIND/01334/2018 (Portugal); the National Priorities Research Program by Qatar National Research Fund; MCIN/AEI/10.13039/501100011033, ERDF ``a way of making Europe", and the Programa Estatal de Fomento de la Investigaci{\'o}n Cient{\'i}fica y T{\'e}cnica de Excelencia Mar\'{\i}a de Maeztu, grant MDM-2017-0765 and Programa Severo Ochoa del Principado de Asturias (Spain); the Chulalongkorn Academic into Its 2nd Century Project Advancement Project, and the National Science, Research and Innovation Fund via the Program Management Unit for Human Resources \& Institutional Development, Research and Innovation, grant B05F650021 (Thailand); the Kavli Foundation; the Nvidia Corporation; the SuperMicro Corporation; the Welch Foundation, contract C-1845; and the Weston Havens Foundation (USA).        
\end{acknowledgments}

\bibliography{auto_generated}

\clearpage
\appendix

\section{Trackless delayed jet tagger input variables} 
\label{TDJInputs}

\begin{figure}[htpb]
        \centering
        \includegraphics[width=0.49\textwidth]{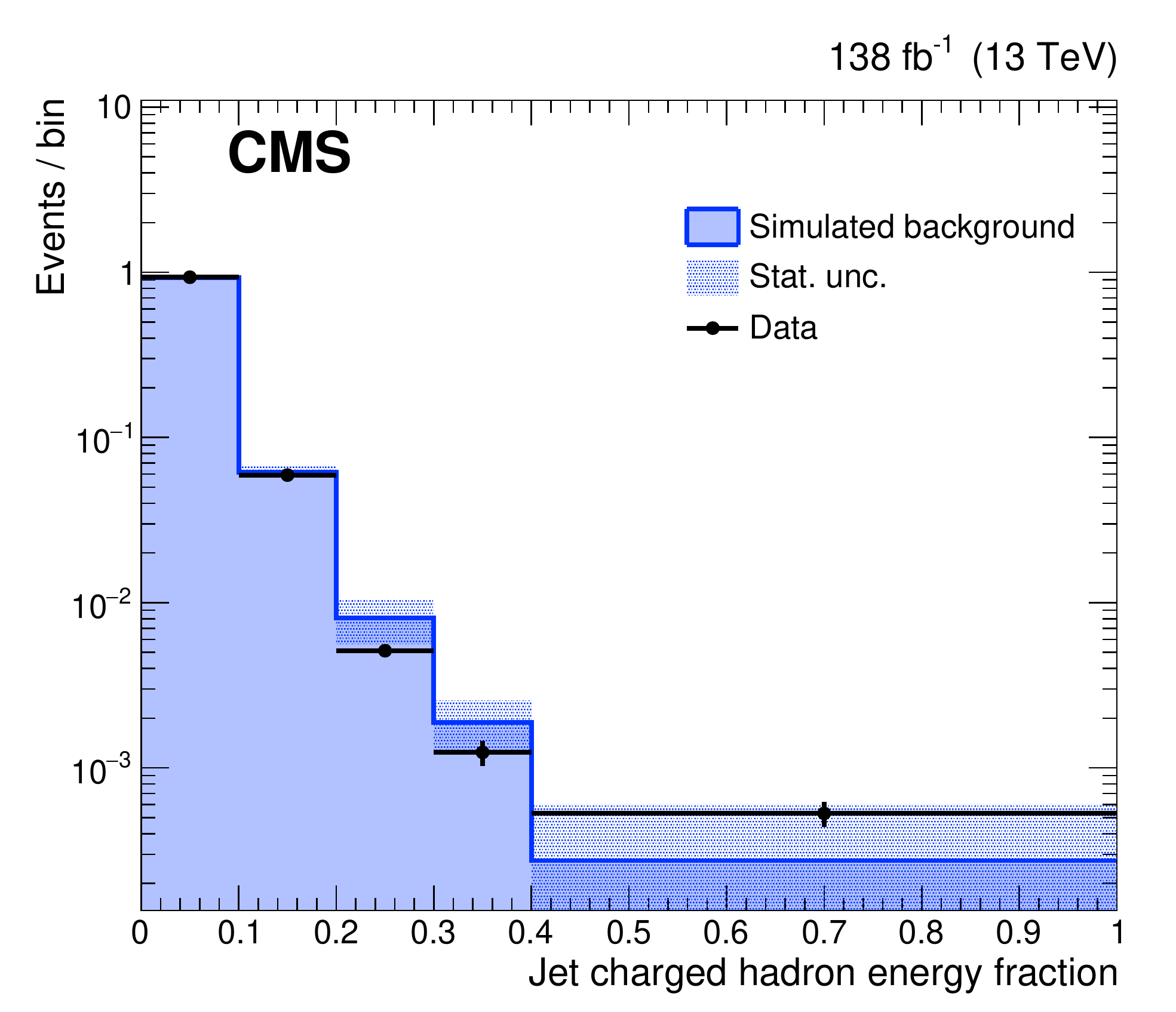}
        \caption{
                The distribution of the jet charged hadron energy fraction, a variable used as input to the TD jet tagger score, for simulation (shaded histogram) and data (black markers) when using electrons from $\PW\to\Pe\Pgne$ events as proxy objects for signal jets. 
                The histograms and data points have been normalized to unit area. 
                Similar levels of agreement are observed for photon proxy objects from the $\PZ\to\Plp\Plm\PGg$ sample.
                \label{fig:TDJ_DataAndBkg_cHadEFrac}
        }
\end{figure}

\begin{figure}[htpb]
        \centering
        \includegraphics[width=0.49\textwidth]{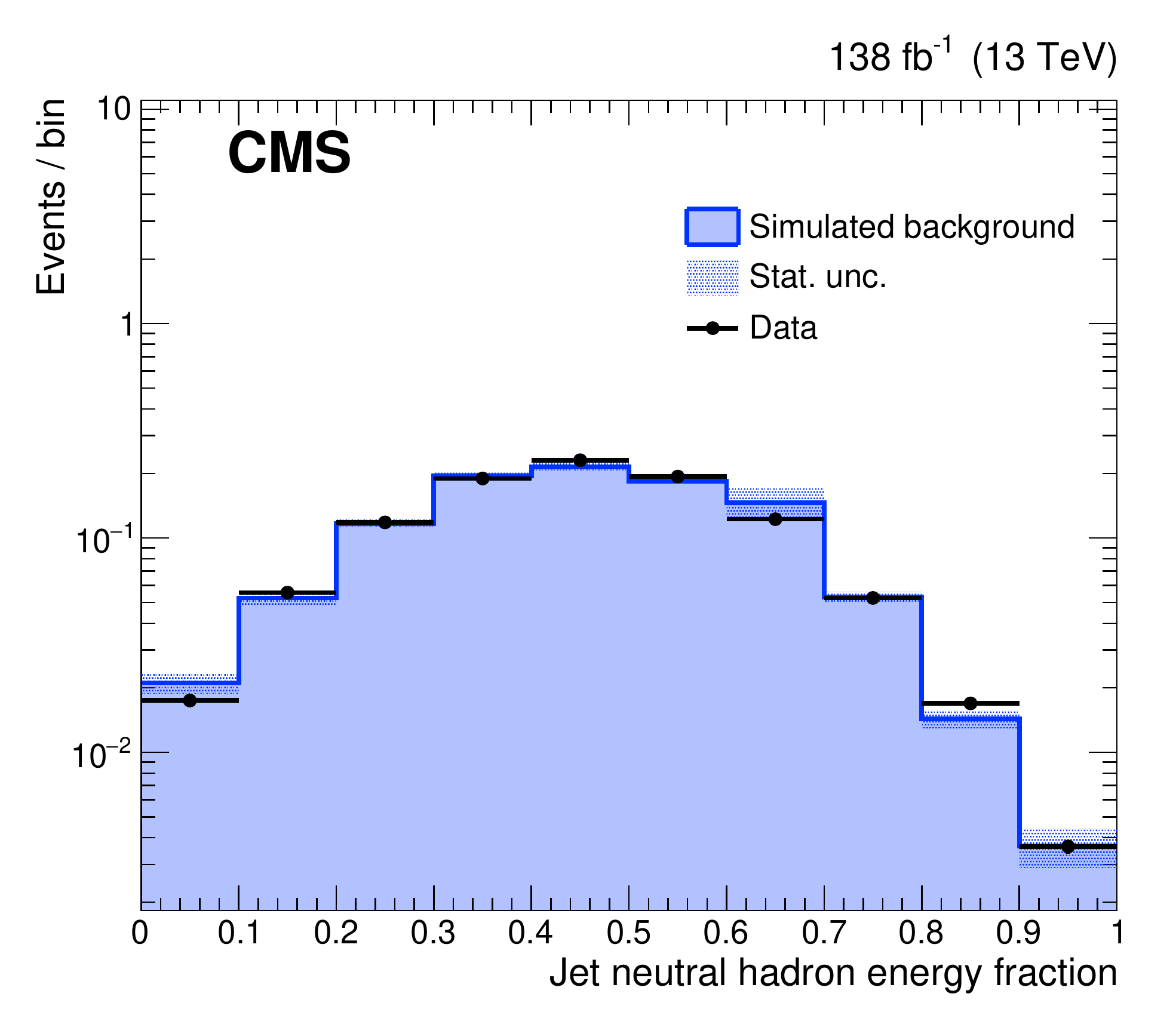}
        \caption{
                The distribution of the jet neutral hadron energy fraction, a variable used as input to the TD jet tagger score, for simulation (shaded histogram) and data (black markers) when using electrons from $\PW\to\Pe\Pgne$ events as proxy objects for signal jets. 
                The histograms and data points have been normalized to unit area. 
                Similar levels of agreement are observed for photon proxy objects from the $\PZ\to\Plp\Plm\PGg$ sample.
                \label{fig:TDJ_DataAndBkg_nHadEFrac}
        }
\end{figure}

\begin{figure}[htpb]
        \centering
        \includegraphics[width=0.49\textwidth]{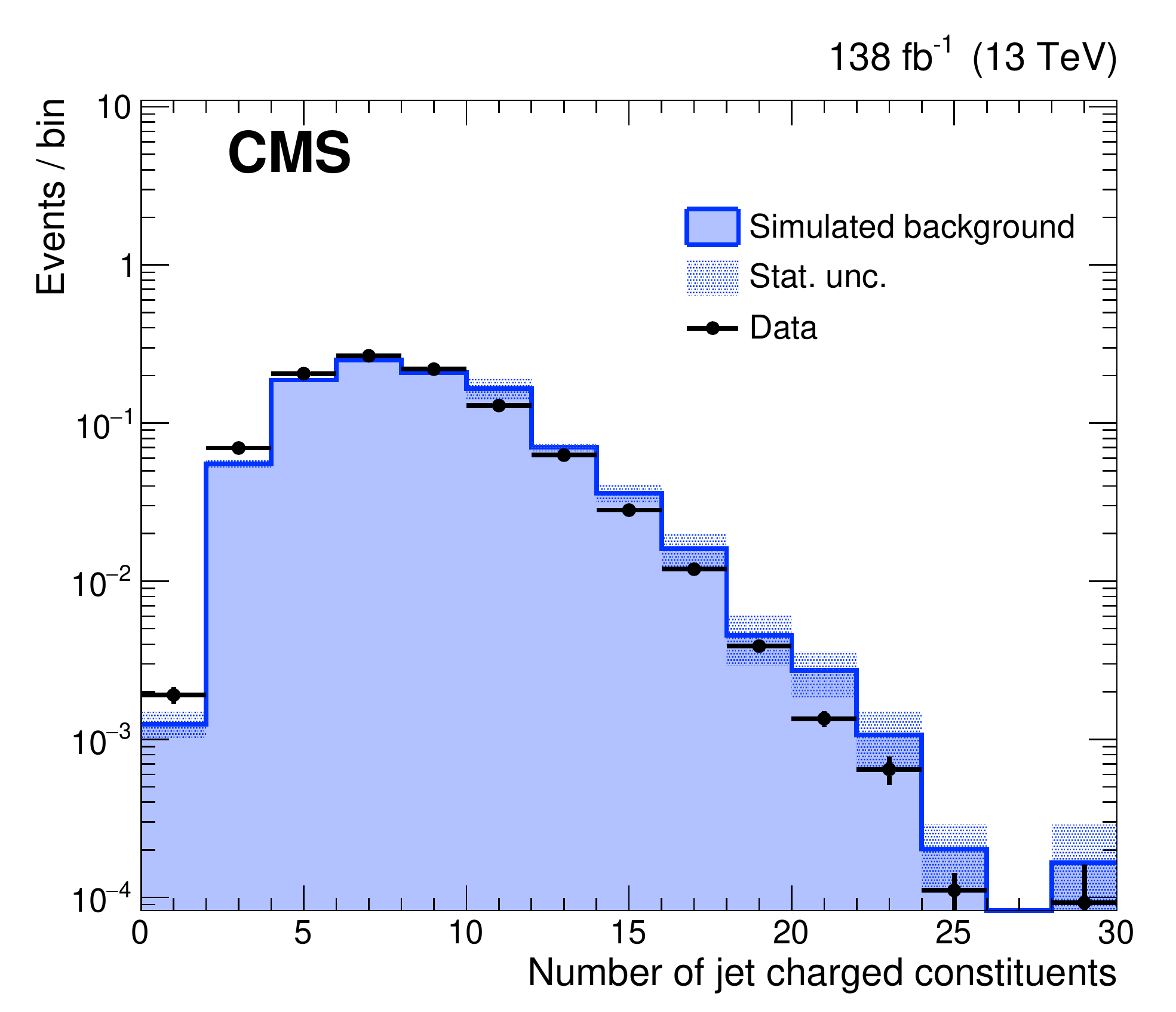}
        \caption{
                The distribution of the number of track constituents in the jet, a variable used as input to the TD jet tagger score, for simulation (shaded histogram) and data (black markers) when using electrons from $\PW\to\Pe\Pgne$ events as proxy objects for signal jets. 
                The histograms and data points have been normalized to unit area. 
                Similar levels of agreement are observed for photon proxy objects from the $\PZ\to\Plp\Plm\PGg$ sample. 
                \label{fig:TDJ_DataAndBkg_nTrackConstituents}
        }
\end{figure}

The key input variables to the TD jet tagger are the charged hadron energy fraction, neutral hadron energy fraction, and number of track constituents.
Comparisons between the simulation-based predictions and data are shown for these three variables in Figs.~\ref{fig:TDJ_DataAndBkg_cHadEFrac},~\ref{fig:TDJ_DataAndBkg_nHadEFrac},~and~\ref{fig:TDJ_DataAndBkg_nTrackConstituents}, respectively. 

\clearpage

\section{TD-tagged jet \texorpdfstring{$\eta$}{eta} distribution}
\label{TaggedJetEta}

\begin{figure}[htpb]
        \centering
        \includegraphics[width=0.49\textwidth]{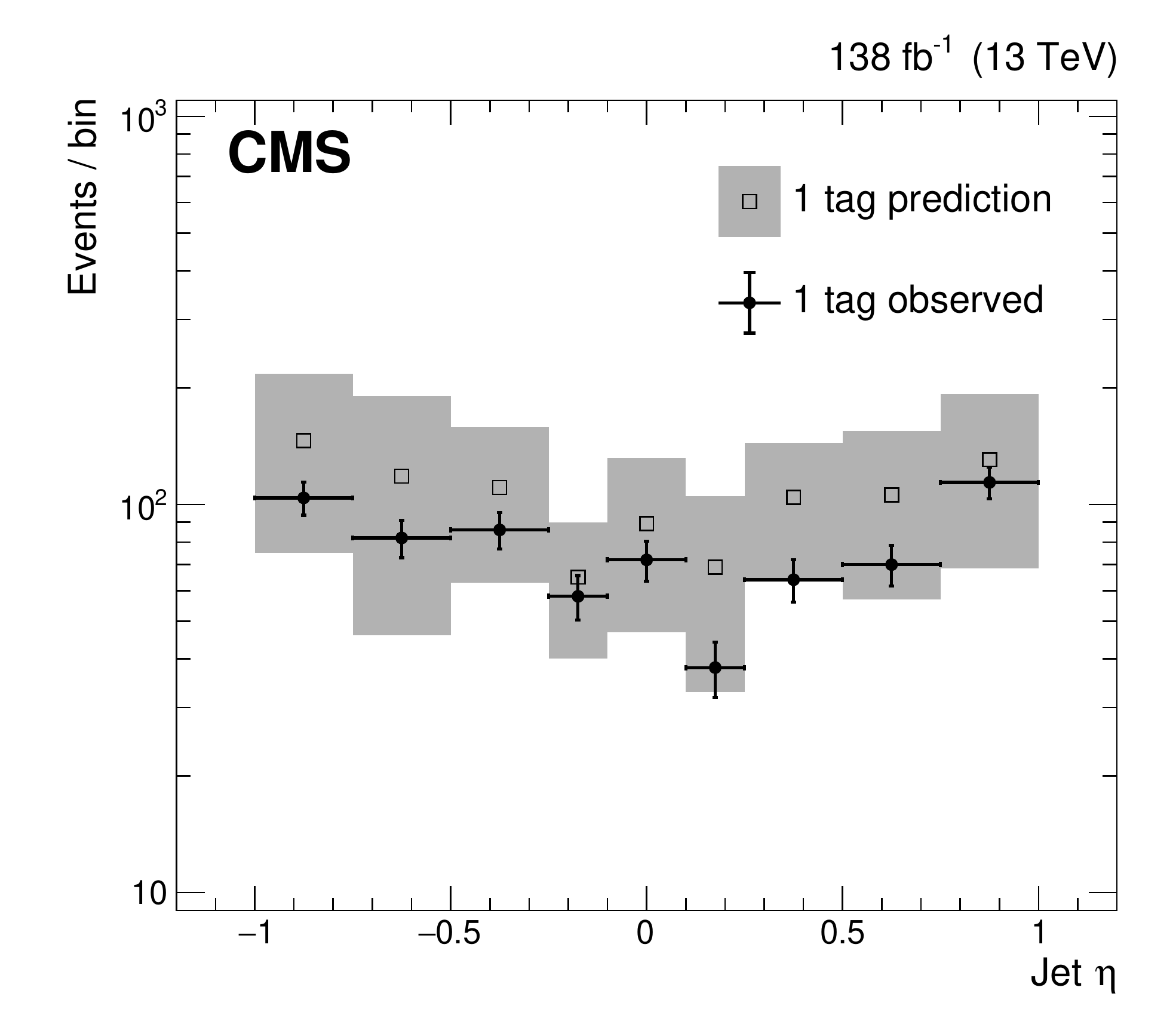}
        \caption{
                The $\eta$ distribution of TD-tagged jets in a background-enriched control region that comprises events with exactly one TD-tagged jet. 
                Observed data (black round markers) and the corresponding prediction based on control samples in data (empty squared markers), measured using the nominal \wjets MR, are compared. 
                The prediction uncertainty (gray band) includes the systematic uncertainty quantifying the degree of process dependence measured from alternative MRs. 
                The predictions for the shape and the normalization of the $\eta$ distribution are consistent with the data.
                \label{fig:TDJ_bin1_prediction}
        }
\end{figure}

The $\eta$ distribution of TD-tagged jets in data and the corresponding prediction from the background estimation method are compared in Fig.~\ref{fig:TDJ_bin1_prediction}.

\cleardoublepage \section{The CMS Collaboration \label{app:collab}}\begin{sloppypar}\hyphenpenalty=5000\widowpenalty=500\clubpenalty=5000
\cmsinstitute{Yerevan Physics Institute, Yerevan, Armenia}
{\tolerance=6000
A.~Tumasyan\cmsAuthorMark{1}\cmsorcid{0009-0000-0684-6742}
\par}
\cmsinstitute{Institut f\"{u}r Hochenergiephysik, Vienna, Austria}
{\tolerance=6000
W.~Adam\cmsorcid{0000-0001-9099-4341}, J.W.~Andrejkovic, T.~Bergauer\cmsorcid{0000-0002-5786-0293}, S.~Chatterjee\cmsorcid{0000-0003-2660-0349}, K.~Damanakis\cmsorcid{0000-0001-5389-2872}, M.~Dragicevic\cmsorcid{0000-0003-1967-6783}, A.~Escalante~Del~Valle\cmsorcid{0000-0002-9702-6359}, P.S.~Hussain\cmsorcid{0000-0002-4825-5278}, M.~Jeitler\cmsAuthorMark{2}\cmsorcid{0000-0002-5141-9560}, N.~Krammer\cmsorcid{0000-0002-0548-0985}, L.~Lechner\cmsorcid{0000-0002-3065-1141}, D.~Liko\cmsorcid{0000-0002-3380-473X}, I.~Mikulec\cmsorcid{0000-0003-0385-2746}, P.~Paulitsch, J.~Schieck\cmsAuthorMark{2}\cmsorcid{0000-0002-1058-8093}, R.~Sch\"{o}fbeck\cmsorcid{0000-0002-2332-8784}, D.~Schwarz\cmsorcid{0000-0002-3821-7331}, M.~Sonawane\cmsorcid{0000-0003-0510-7010}, S.~Templ\cmsorcid{0000-0003-3137-5692}, W.~Waltenberger\cmsorcid{0000-0002-6215-7228}, C.-E.~Wulz\cmsAuthorMark{2}\cmsorcid{0000-0001-9226-5812}
% ...
\par}
\vskip\cmsinstskip
\dag:~Deceased\\
$^{1}$Also at Yerevan State University, Yerevan, Armenia\\
$^{2}$Also at TU Wien, Vienna, Austria\\
% ...
\end{sloppypar}
\end{document}